\documentclass[11pt]{article}

\usepackage{cite}

\usepackage{epsfig}

\usepackage{amsmath}
\usepackage{amssymb}

\begin{document}
%

%
%

\def\lap{\nabla^2}

\def\gtt{G^{t}_{~t}} \def\grr{G^{r}_{~r}} \def\grz{G^{r}_{~z}}
\def\gfg{G^{\rho}_{~\chi}} \def\gzz{G^{z}_{~z}}
\def\gaa{G^{\theta}_{~\theta}} \def\grrzz{(G^{r}_{~r}+G^{z}_{~z})}
\def\grrzz{(G^{\rho}_{~\rho}+G^{\chi}_{~\chi})}
\def\crz{G^{r}_{~z}} \def\crrzz{(G^{r}_{~r}-G^{z}_{~z})}
\def\gffgg{(G^{\rho}_{~\rho}+G^{\chi}_{~\chi})}
\def\cfg{G^{\rho}_{~\chi}}
\def\cffgg{(G^{\rho}_{~\rho}-G^{\chi}_{~\chi})}

\def\figuremode{\small}

%
%

\title{\bf Properties of Kaluza-Klein black holes}

\author{ {\bf Hideaki Kudoh}\thanks{e-mail: {\tt
      kudoh@yukawa.kyoto-u.ac.jp}} \\ \\
  Department of Physics, \\
  Kyoto University, \\
  Kyoto 606-8502, Japan
  \\ \\
  {\bf Toby Wiseman}\thanks{e-mail: {\tt
      twiseman@fas.harvard.edu}}
 \\ \\
Jefferson Physical Laboratory \\
Harvard University \\
Cambridge, MA 02138, USA }

\date{October 2003}

\maketitle

%
\begin{abstract}
%
  
  We detail numerical methods to compute the geometry of static vacuum
  black holes in 6 dimensional gravity compactified on a circle. We
  calculate properties of these Kaluza-Klein black holes for varying
  mass, while keeping the asymptotic compactification radius fixed.
  For increasing mass the horizon deforms to a prolate ellipsoid, and
  the geometry near the horizon and axis decompactifies. We are able
  to find solutions with horizon radii approximately equal to the
  asymptotic compactification radius.  Having chosen 6-dimensions, we
  may compare these solutions to the non-uniform strings compactified
  on the same radius of circle found in previous numerical work. We
  find the black holes achieve larger masses and horizon volumes than
  the most non-uniform strings.  This sheds doubt on whether these
  solution branches can merge via a topology changing solution.
  Further work is required to resolve whether there is a maximum mass
  for the black holes, or whether the mass can become arbitrarily
  large.

%
\end{abstract}
%

\vspace{3.0cm}
\begin{flushright}
hep-th/0310104 \\
DAMTP-2003-103\\
KUNS-1872
\end{flushright}

\newpage


\newpage

%
\section{Introduction}
\label{sec:intro}
%

If one day we discover that there are extra dimensions in our
universe, and these are well described by classical gravity, then they
are likely (although not definitely \cite{Randall_Sundrum2}) to be
compact along the lines of Kaluza-Klein theory \cite{Kaluza,Klein}.
Furthermore, if matter is confined to branes, then the radius of
compactification could potentially be extremely large \cite{ADD,AADD}.  The
simplest regular static vacuum solutions are then compactified uniform
black strings \cite{Gibbons_Wiltshire,Chodos}.  Gregory and Laflamme
(GL) showed that these are stable provided the horizon radius is large
compared to the compactification scale
\cite{Gregory_Laflamme2,Gregory_Laflamme1,Gregory_Laflamme3}.
However, they also discovered a new family of non-uniform solutions
emerging from the critical uniform string whose mass separates the
more massive stable strings from the less massive unstable ones. These
non-uniform solutions were constructed numerically, firstly by Gubser
as a perturbation expansion in a non-uniformity parameter $\lambda$
about the $\lambda = 0$ critical uniform solution \cite{Gubser}, and
then non-perturbatively in \cite{Wiseman3} using elliptic methods.
The third class of solutions expected to exist are black holes that do
not wrap the circle direction.  In 4 dimensions such solutions were
found by Myers analytically \cite{Myers} (see also generalisations
\cite{Nicolai, Frolov} with modified asymptotics) but in more than 4
dimensions little is known \cite{Emparan_Reall1,Harmark_Obers},
essentially as the rotation group then has curvature.  Using the
elliptic numerical methods of \cite{Wiseman1, Wiseman3} 5 dimensional
localised black holes have recently been constructed on a
Randall-Sundrum brane \cite{Kudoh1,Kudoh2}. This is a related
numerical problem of considerable interest as recent conjectures claim
large localised static black holes may not exist
\cite{Tanaka,Emparan_Kaloper}.  

Kol proposed an elegant relation between these three types of
Kaluza-Klein solution \cite{Kol1}, the non-uniform strings linking the
uniform branch to the black hole branch. The string to black hole
transition, also explored in
\cite{Harmark_Obers,Kol2,Piran,Harmark_Obers2,Kol_Piran2,Harmark_Obers3},
is then conjectured to be continuous, and have a Lorentzian cone
geometry where the horizon degenerates, and this agrees very well with
numerical tests on the non-uniform string branch
\cite{Wiseman4,Kol_Wiseman}.  If this picture is correct, it predicts
that the black holes, like the non-uniform strings, have a maximum
mass.  While for fixed compactification radius the 3 classes of
solution overlap at intermediate mass scales, \footnote{We must
  consider the full asymptotic charges to distinguish the solution
  \cite{Harmark_Obers2,Kol_Piran2}, although whether there is a unique
  solution with these charges is an interesting open question
  \cite{Harmark_Obers3}, as uniqueness constraints apparently weaken
  in more than 4 dimensions \cite{Emparan_Reall2,Kol3}.} at large
masses the uniform strings would be the unique non-singular solutions
in Kaluza-Klein theory.  If incorrect, it may be possible to have
arbitrarily high mass black holes, if either the geometry becomes
increasingly `squashed' or decompactifies on the symmetry axis (as in
pure Kaluza-Klein theory there is no radius stabilisation). Whether
this could persist in a radius stabilised theory would then be an
important phenomenological question.

Clearly Kaluza-Klein theory is a simplification of realistic
compactifications. Kol's picture presumably remains unchanged adding
warping \cite{Randall_Sundrum1,Reall2,Gregory,Gibbons_Hartnoll1}, or
charging black holes under matter fields localised to branes
\cite{Kanno_Soda2}. For additional bulk matter, such as is necessary
for stabilisation, the situation may be more interesting, but we
expect it will inherit many features of the pure Kaluza-Klein case.

The Gregory-Laflamme instability, underlying the dynamics of these
compactified horizons has been linked to thermodynamic stability
\cite{Gubser_Mitra1,Gubser_Mitra2,Reall,Ross,Hubeny_Rangamani,Gubser_brane,Gibbons_Hartnoll1,Gibbons_Hartnoll2,Hartnoll2},
and an analogous classical instability has recently been conjectured
for the rotating Myers-Perry solution
\cite{Myers_Perry,Reall_susybh,Emparan_Myers}, which is thought to be
unstable for large angular momenta. The end-state of the classical
Gregory-Laflamme instability is still a mystery, although there has
been interesting analytic and numerical work on this subject
\cite{Horowitz, Choptuik}. We note that stable black strings
evaporating via Hawking radiation will eventually succumb to this
classical instability, and understanding the dynamics, and in
particular whether cosmic censorship is violated, is important in
order to understand evaporation of cosmological/astrophysical black
holes below the compactification mass scale.

The objective of this paper is to numerically construct and study the
non-wrapping black hole branch of solutions. We perform this analysis
in 6-dimensions so that we may compare with the previous non-uniform
string numerical results of \cite{Wiseman3}, which for technical
reasons, were performed in this number of dimensions. We begin with a
brief discussion of the numerical method, which involves phrasing a
subset of the Einstein equations in a way compatible with numerical
relaxation, and most importantly, showing how the remaining
`constraint' equations can be satisfied by appropriately choosing the
boundary conditions.  Since the method has now been used several times
\cite{Wiseman1, Wiseman3, Kudoh1} we refrain from a detailed
exposition, and instead highlight the various subtleties related to
this Kaluza-Klein black hole problem. We then go on to discuss the
numerical results. We demonstrate that as expected these solutions
exist, at least within the scope of our numerical approximation.

We compute geometric embeddings of the spatial horizon and symmetry
axis into Euclidean space, and show that the geometry near the axis
decompactifies with increasing mass, and the event horizon deforms to
a prolate ellipsoid.  With the current implementation we are unable to
ascertain whether this decompactification terminates with a maximum
mass black hole that just `fits' into the compact direction, or
whether the decompactification continues indefinitely so that
arbitrarily high masses can be found. 

The maximum size black holes we are able to construct have horizon
radii approximately equal to the asymptotic compactification radius.
We compare these with the most non-uniform strings constructed in
\cite{Wiseman3} finding the mass and horizon volume of these modest
sized black holes already becomes larger than that of the maximally
non-uniform strings, and the axis decompactifies to a greater extent.
The size of black hole we may construct is limited by numerical
factors, and it seems clear that still larger black holes exist, with
the above trends continuing for these. The implication is that it
appears \emph{unlikely} that the non-uniform string branch (connected
to the critical uniform string) and \emph{this} black hole branch are
connected via a topology changing solution.

Various technical details and numerical checks are reserved for the
three Appendices. We pay particular attention to ensuring and checking
that the constraint equations are indeed satisfied for the solutions.

The reader is also referred to independent work by Kol, Piran and
Sorkin who we understand have recently performed related calculations
in 5 dimensions \cite{Kol_Piran3}.

%
\section{Method}
\label{sec:method}
%

In order to solve the black hole geometry we are required to solve the
Einstein equations with elliptic boundary data; we wish to have a
regular horizon geometry, for the solution to be periodic, and also
asymptotically to tend to flat space product with a circle. We
employ the methods first developed in \cite{Wiseman1} and used to
construct non-uniform string solutions \cite{Wiseman3}, and later
localised black holes on branes\cite{Kudoh1,Kudoh2}.
\footnote{See also \cite{Kunz} for a method of solving static
  axisymmetric \emph{non-vacuum} black holes in 4-dimensions which
  shares some features with our method.}
In this section we outline the method and
boundary conditions appropriate for the problem.  Due to its
necessarily technical nature, some readers may wish to skip to the
following `Results' section. For a more general discussion of the
method, the reader is referred to \cite{Wiseman3}. Technical
numerical details are also provided for the interested reader in
Appendix \ref{app:details}, and important numerical checks are
reported in Appendix \ref{app:checks} to demonstrate the method
performs correctly.

Constructing the non-uniform strings is a very clean situation in
which to apply these elliptic numerical methods. However, the black
hole problem at hand is substantially more difficult, primarily for
two reasons;

Firstly weakly non-uniform strings can be described as a perturbative
deformation of the critical uniform string. The relaxation methods
employed here require a good initial guess, or typically no solution
will be found.  Thus for the strings the non-uniformity can be turned
on `gently'.  In analogy, a very small black hole will appear as a 6-d
Schwarzschild solution near its horizon, but it obviously must have
very different asymptotics due to the compactification. Thus even for
a small black hole, we do not have an exact solution to `gently' start
building larger black holes from. We tackle this issue by building in
6-d Schwarzschild behaviour at the horizon that decays quickly away
far from the horizon, and then we solve for the correction to this,
which should be small for low mass black holes.

Secondly, the axis of symmetry is exposed in the problem.  The
coordinate singularity at the axis generically gives rise to problems
numerically and there are various ways around this in conventional
evolution problems (see for example \cite{Choptuik_axisym}). As
discussed in \cite{Wiseman1} the elliptic method we use is very
sensitive to this coordinate singularity, which may destroy the
ability of the algorithm to relax to the solution. Furthermore the
coordinate system we require to phrase a subset of the Einstein
equations in an elliptic manner introduces even worse coordinate
problems on the symmetry axis than one would normally expect. We have
found no elegant method to tackle this problem, but do have a
functional approach, originally used in \cite{Wiseman1} and discussed
here in Appendix \ref{app:details}.  Improving or evading this problem
appears to be crucial for increasing the capability of this method.

So whilst the problem is a rather delicate one, we are still able to
make progress. As in previous applications of the method, we write the
static axisymmetric metric in a diagonal form, retaining a conformal
invariance in the radial and tangential coordinate $r,z$ as;
\begin{equation}
ds^2 = g_{MN} dx^M dx^N= - e^{2 \alpha} dt^2 + e^{2 ( \beta - \gamma )} \left( dr^2 + dz^2 \right) + r^2 e^{2 \beta + \frac{4}{3}\gamma} d\Omega^2_{(3)}
\label{eq:simplemetric}
\end{equation}
with $\alpha, \beta, \gamma$ being functions of $r, z$. The particular
linear combination of $\beta, \gamma$ taken above is simply for
technical convenience later. We take the $z$ coordinate to be compact
with period $L$, and later will require the metric functions $\alpha,
\beta, \gamma$ to vanish at large $r$, and hence the physical radius
of compactification will be $L$. We choose units such that the
6-dimensional Newton constant, $G_{N(6)} = 1$.  Since we may perform a
global scaling on any solution of the vacuum Einstein equations, for
future convenience we chose to set $L = \pi$ in these units.

One nice property of this form of the metric is that one can choose
the position of the boundaries in the $(r,z)$ plane to be at any
location, due to the residual conformal coordinate transformations. A
second important feature of this coordinate system is that 3 of the 5
Einstein equations, $\gtt, \grrzz, \gaa$ have elliptic second
derivatives, being just the $(r,z)$ Laplace operators, and thus we
term these the `elliptic' equations.  As we are so far unable to write
a positive definite functional of the metric components which can be
minimised to give these equations, it is not at all clear the problem
is truly elliptic.  However one can still use relaxation methods to
solve them, specifying elliptic data on the boundaries of the problem.
The most important feature of this coordinate system is that using the
contracted Bianchi identities, the 2 remaining `constraint' equations
weighted by $\det{g_{MN}}$,
\begin{equation}
\hat{\Phi} = \det{g_{MN}} \crz \qquad \hat{\Psi} = \frac{1}{2} \det{g_{MN}} \crrzz
\end{equation}
obey Cauchy-Riemann (CR) relations,
\begin{equation}
\partial_r \hat{\Phi} = \partial_z \hat{\Psi} \qquad \qquad \partial_z \hat{\Phi} = - \partial_r \hat{\Psi}
\end{equation}
if the elliptic equations are satisfied.  Simply relaxing the 3
elliptic equations for the 3 metric functions will generically yield a
solution, but this is \emph{only} consistent with the full set of
Einstein equations if the boundary data is such that the CR relations
imply both constraints are satisfied.  For example, for the
non-uniform strings we may impose the constraints by updating the
elliptic boundary data such that the $\crz$ constraint is satisfied on
all boundaries, and the remaining constraint $\crrzz$ is just imposed
at one point, solving the CR problem \cite{Wiseman3}. In our example
here, we in fact find it convenient to impose both constraints, but on
different boundaries, in such a way that still provides sufficient
conditions to satisfy the CR problem, but does not over determine the
elliptic equations.

\subsection{Boundaries and coordinates}

We still have residual coordinate freedom, and we use this to tailor
the coordinate system to our problem. Instead of relaxing $\alpha,
\beta, \gamma$ directly, we wish to perturb about a 6-d Schwarzschild
solution near the horizon,
\begin{equation}
\alpha = A_{bg} + A \qquad 
\beta  = B_{bg} + B \qquad
\gamma = C_{bg} + C 
\end{equation}
where the `background functions' $\{A,B,C\}_{bg}$ suitably express
this Schwarzschild geometry at the horizon, decay away radially, and
are compatible with the compact boundary conditions - so we could
obviously not just take the Schwarzschild metric itself. Notice that
$C = 0$ corresponds to conformally flat spatial sections.

\begin{figure}
\centerline{\psfig{file=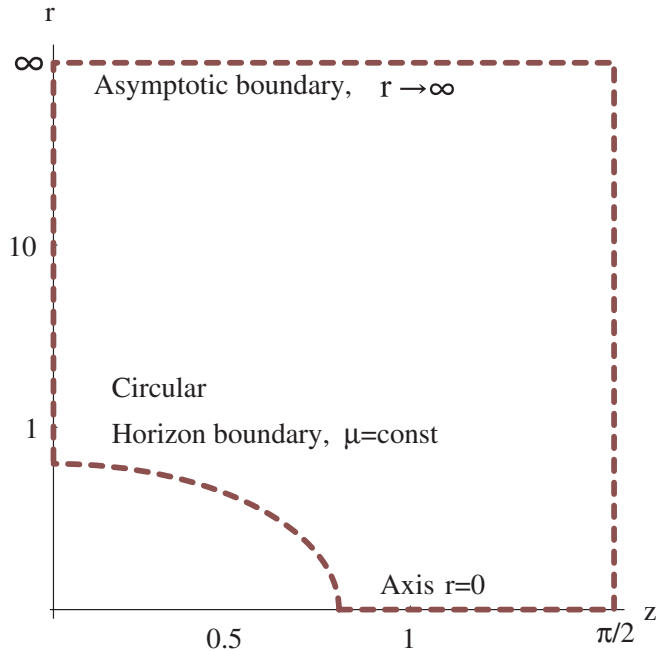,width=3.4in}}
\caption{
\label{fig:rzboundaries}
Schematic illustration of the boundaries we would intuitively take in
$r,z$ coordinates. At least for small black holes, taking the horizon
boundary to be circular would ensure the metric functions $A,B,C$
would be small and finite, as the geometry near the horizon would only
be a weak distortion of 6-dimensional Schwarzschild, which we build
into the metric ansatz via $A_{bg},B_{bg},C_{bg}$.  }
\end{figure}

Let us now consider a form for the background functions, and boundary
locations in the $(r,z)$ coordinates. The 6-d Schwarzschild metric can
easily be written in an appropriate conformal form as,
\begin{eqnarray}
\label{eq:schwarz}
ds^2 & = & - n(\mu) dt^2 + a(\mu) \left( d\mu^2 + \mu^2 d\Omega^2_{(4)} \right)
\\ \nonumber 
& = & - \left[\frac{\mu^3-\mu_0^3}{\mu^3+\mu_0^3}\right]^2 dt^2 + \left[ 1 + \frac{\mu_0^3}{\mu^3}\right]^{\frac{4}{3}} \left( dr^2 + dz^2 + r^2 d\Omega^2_{(3)} \right)
\end{eqnarray}
with $\mu^2 = r^2 + z^2$. To be suitable for the background functions,
this must be modified away from the horizon to ensure compatibility
with the periodic boundary conditions in $z$.  However, assuming after
this modification the horizon remains at constant $\mu = \mu_0$, it
will form a circle in the $(r,z)$ coordinates. Thus we would wish to
take boundaries of the form in figure \ref{fig:rzboundaries} in order
to represent the coordinate axis and black hole horizon, and periodic
boundaries. Then we would reasonably expect the functions $A, B, C$ to
remain finite everywhere, and be small for a small black hole,
allowing us to use the initial data $A = B = C=0$ for the relaxation.

\begin{figure}
\centerline{\psfig{file=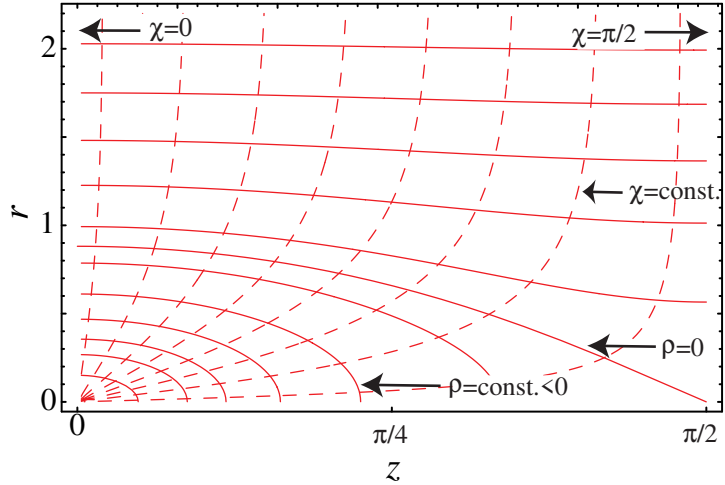,width=4in}}
\caption{
  Illustration of isosurfaces of $\rho, \chi$ as functions of $r,z$.
  The schematic boundaries of figure \ref{fig:rzboundaries} can be
  mapped to constant $\rho$ and $\chi$ values, making for convenient
  numerical implementation.  For $\rho<0$, $\chi$ behaves as an
  angular coordinate, and $\rho$ as a radial one, whereas for large
  positive $\rho$, we find $\rho, \chi$ behave as $r, z$ respectively.
  We will take the horizon boundary at constant $\rho = \rho_0$,
  finding later that specifying $\rho_0$ determines the physical size
  of the black hole solution.
\label{fig:fgboundaries}
}
\end{figure}

Numerically it is always convenient to have a rectangular grid. Whilst
we are in principle free to use the residual conformal transformation
to fix the boundaries to be wherever we wish, clearly to obtain a
rectangular domain such a transformation must be singular, as one
right-angle in the figure \ref{fig:rzboundaries} should be `flattened'
out. However, if we find such a coordinate transformation
analytically, we may separate out any singular behaviour from $A, B,
C$, leaving their behaviour perfectly regular. Any conformal
coordinate transformation is generated by a solution to the 2-d
Laplace equation, and choosing a solution $\rho(r,z)$ to be that
representing a point source in the compact 2-d space (ie.  on a
cylinder) we may define new coordinates $(\rho,\chi)$,
\begin{eqnarray}
\rho(r,z) & = & \frac{1}{2} \log \left[ \frac{1}{2} \left( \cosh{2 \, r} - \cos{2\,z} \right) \right]
\\ \nonumber
\chi(r,z) & = & \tan^{-1} \left[ \frac{\tan z}{\tanh r} \right]
\end{eqnarray}
where $\chi$ is determined from $\rho$ by CR relations. These
essentially `flatten' out the horizon, and now $\chi$ is the compact
coordinate which conveniently takes the range of an angular coordinate
$(0, \pi/2)$ for half a period of the solution, $z = (0,L/2) =
(0,\pi/2)$. We illustrate the isosurfaces of $\rho$ and $\chi$ in the
$(r,z)$ plane in figure \ref{fig:fgboundaries}. Contours of constant
$\rho$, for $\rho<0$, generate very similar curves in the $(r,z)$
plane to that of the horizon in figure \ref{fig:rzboundaries}.  Note
also that $\chi = \pi/2$ gives us \emph{both} the axis of symmetry
(for $\rho<0$) \emph{and} the periodic boundary $z = \pi/2$ (for
$\rho>0$), and $\rho = 0$, $\chi = \pi/2$ is the singular point in the
conformal transformation.  Thus if we use these coordinates, a
rectangle with $\chi = 0$ to $\pi/2$ and $\rho = \rho_0$ to $\infty$
will, for $\rho_0<0$, give us similar looking boundaries to figure
\ref{fig:rzboundaries}.

Whilst the coordinate transform is rather singular, since we have an
analytic expression for it, we may remove the singular Jacobian, $J$,
and now write the metric,
\begin{eqnarray}
\label{eq:jacmetric}
ds^2 =  - e^{2 \alpha} dt^2 & + & e^{2 (\beta - \gamma)} J(\rho,\chi) \left( d\rho^2 + d\chi^2 \right) + r(\rho,\chi)^2 e^{2 \beta + \frac{4}{3} \gamma} d\Omega^2_{(3)}
\\ \nonumber
\\ \nonumber
J(\rho,\chi) & = & \frac{e^{2 \rho}}{\sqrt{1 + e^{4 \rho} + 2 \, e^{2 \rho} \cos{2 \chi}}}
\\ \nonumber
\\ \nonumber
r(\rho,\chi) & = & \frac{1}{2} \cosh^{-1}\left[ e^{2 \rho} + \sqrt{1 + e^{4 \rho} + 2 \, e^{2 \rho} \cos{2 \chi}} \right]
\end{eqnarray}
where now $\alpha, \beta, \gamma$ are exactly the same functions as
previously in \eqref{eq:simplemetric}, except now in terms of $\rho,
\chi$.

Now consider the 6-d Schwarzschild metric \eqref{eq:schwarz}. For
small $\mu_0$, near the horizon we find $e^{2 \rho} \simeq \mu^2 = r^2
+ z^2$, with $\rho$ behaving as a polar radial coordinate and $\chi$
as the angular one. Since we wish to choose background functions to
reproduce the Schwarzschild metric near the horizon for small black
holes (ie. small $\mu_0$), and yet to implement the compactness
requirement in $z$, a simple choice is just to substitute $e^{\rho -
  \rho_0} = \mu/\mu_0$ into \eqref{eq:schwarz} giving,
\begin{eqnarray}
e^{A_{bg}} & = &  \frac{1 - e^{3 \left( \rho_{0} - \rho \right) }}{1 + e^{3 \left( \rho_{0} - \rho \right) }}
\\ \nonumber
e^{B_{bg}} & = &  \left[ 1 + e^{3 \left( \rho_{0} - \rho \right) } \right]^{\frac{2}{3}}
\\ \nonumber \\ \nonumber
C_{bg} & = & 0
\end{eqnarray}
The independence of these background functions on $\chi$ ensures they
satisfy the periodicity requirements.  The constant $\rho_0$ tells us
the coordinate position of the horizon.  We will shortly show that
keeping the asymptotic compactification radius fixed, it is the
parameter $\rho_0$, and thus the contour $\rho = \rho_0$ we take to be
the horizon that we will vary to change the mass of the black hole.
From the earlier figure \ref{fig:fgboundaries} we see that $\rho_0$
must be negative to give a spherical topology for the horizon.  The
small horizon limit is now $\rho_0 \rightarrow - \infty$. For
$\rho_0<0$ but closer to zero, the ansatz above gives a deformed
geometry from Schwarzschild too. 
\footnote{It is interesting to compare this coordinate system with
  that proposed by Harmark and Obers in \cite{Harmark_Obers} where the
  radial coordinate used followed a 6-d equipotential. Here, our
  coordinate follows a 2-d equipotential. Presumably using the higher
  dimensional potential is a sensible procedure, and could give an
  improved ansatz to perturb about, which better models the horizon
  geometry for the larger deformed black holes. It remains an
  interesting problem where one can use the ansatz of Harmark and
  Obers to do numerics with, particularly as the ansatz and having the
  horizon at constant potential location was proven to be consistent
  for non-uniform strings in \cite{Wiseman4} and recently for black
  holes in \cite{Harmark_Obers3}}
Since we include the vanishing of the lapse in $A_{bg}$ at
$\rho=\rho_0$ we expect the metric functions, and in particular $A$,
the one associated with deformations of the lapse, to remain finite
there. Far away from the horizon, ie. for $\rho \rightarrow \infty$ or
alternatively $r \rightarrow \infty$, the background functions decay
exponentially in $\rho$ or $r$.  We stress that for finite negative
$\rho_0$, $A, B, C = 0$ is \emph{not} a solution to the Einstein
equations, but for very negative $\rho$, the small black hole limit,
$A, B, C$ will at least be small everywhere as the horizon tends to
Schwarzschild, and by the time the metric functions `see' the
compactification they will be vanishingly close to zero anyway.  After
the change in coordinates, we obtain analogous elliptic equations,
$\gtt, \gffgg, \gaa$ and also CR relations for the new constraint
functions $\Phi, \Psi$,
\begin{equation}
\Phi = \det{g_{MN}} \cfg \qquad \Psi = \frac{1}{2} \det{g_{MN}} \cffgg 
\end{equation}
so that,
\begin{equation}
\partial_\rho \Phi = \partial_\chi \Psi \qquad \qquad \partial_\chi \Phi = - \partial_\rho \Psi
\end{equation}

\subsection{Boundary conditions from the elliptic equations}

The 3 elliptic equations we are solving require various boundary
conditions due to the regular singular or periodic behaviour at these
boundaries.  To satisfy the constraint equations we must impose more
than just these conditions. However, let us start by considering the
basic boundary conditions from the elliptic equations.

Asymptotically we want the geometry to be a product of 5-dimensional
flat space with a circle, and thus we take $A,B,C \rightarrow 0$.
Since we earlier fixed the range of $\chi$, this also fixes the
compactification radius to be $L = \pi$. Of course we may simply
globally scale these vacuum solutions to obtain any desired asymptotic
radius.  We find the asymptotic form required by the 3 elliptic
equations,
\begin{eqnarray}
\label{eq:asym}
A \sim \frac{a_2}{\rho^2} + O\left(\frac{1}{\rho^3}\right) \qquad B \sim \frac{b_1}{\rho} + \frac{b_2}{\rho^2}  + O\left(\frac{1}{\rho^3}\right) \qquad C & \sim & \frac{b_1}{\rho} + \frac{c_2}{\rho^2} + O\left(\frac{1}{\rho^3}\right) 
\\ \nonumber \\ \nonumber
\mbox{with} \qquad 3 \, a_2 + 9 \, b_2 - 4 \, c_2 + \frac{25}{6} b_1^2 & = & 0
\end{eqnarray}
expanding (without linearising in the metric components) in inverse
powers of $\rho \sim r$. Fourier modes with $\chi$ dependence decay
exponentially (since $\chi \simeq z$, for large $\rho$), as does the
contribution from the background functions $\{A,B,C\}_{bg}$.
Thus on our asymptotic boundary we have mixed Neumann-Dirichlet
boundary conditions for the 3 metric functions.
\footnote{ In practice we impose these conditions at a finite, but
  large $\rho = \rho_{max}$ and check in Appendix \ref{app:checks}
  that the results are independent of $\rho_{max}$.  }

The symmetry axis $\chi = \pi/2$ with $\rho<0$ requires that $A, B$ be
even in $(\chi-\pi/2)$ (or alternatively $r$). However we also find
the requirement that $C = 0$ as there is a regular singular behaviour
due to the form of the coordinate system. We might be confused that
there is no Neumann condition on $C$ but we see later this emerges
from the constraints.

We require that the metric functions $A,B,C$ be finite at the horizon
$\rho = \rho_0$.  With the form of background functions
$\{A,B,C\}_{bg}$ we then find that the elliptic equations are
consistent provided that,
\begin{eqnarray}
\label{eq:horizonbc}
A_{,\rho} \mid_{\rho = \rho_0} & = & 0
\\ \nonumber
\\ \nonumber
B_{,\rho} + \frac{2}{3} C_{,\rho} \mid_{\rho = \rho_0}& = & 1 - \frac{r_{,\rho}}{r} \mid_{\rho = \rho_0}
\end{eqnarray}
with $r(\rho,\chi)$ as given earlier in equation \eqref{eq:jacmetric}.

Finally, the elliptic equations at the remaining periodic boundaries
at $\chi = 0$ and $\chi = \pi/2$ with $\rho>0$ simply imply Neumann
conditions on the metric functions $A,B,C$.

\subsection{Boundary conditions from the constraints}

The constraint equations also impose conditions on the metric. We have
seen that assuming we can satisfy the elliptic equations for given
boundary data, the two constraints obey CR relations.  Thus we do not
need to enforce both constraints on all the boundaries, which naively
would over-determine the elliptic equations.  Firstly we consider the
extra conditions the constraints impose, and then discuss how best to
implement them to ensure a consistent solution of the CR problem,
without over determining the elliptic data.

On the symmetry axis and periodic boundaries we have already specified
3 conditions, one for each metric function, and consequently treating
this as a boundary value problem, we do not wish to impose any more.
On the $\chi = 0$ periodic boundary $\cfg$ vanishes by symmetry, and
consequently so does the corresponding weighted constraint $\Phi$, and
the remaining constraint $\cffgg$ is guaranteed to be even. A similar
situation occurs on the $\chi = \pi/2$ boundary for $\rho>0$ which
represents the other periodic boundary.  Indeed these periodic
boundaries are fictitious in the sense that we can consider the
problem on the unwrapped covering space where these boundaries are
`removed', and thus we should not need to impose any constraints here.

For $\chi = \pi/2$, but now with $\rho<0$ we have the symmetry axis.
Again this is in principle a `fictitious' boundary, but we must impose
$C = 0$ here for the elliptic equations, and it is hard to see how
this would lift to a covering space with `no' boundary. Thus we
examine the situation at the axis in more detail. Using the boundary
conditions from the elliptic equations, and assuming the metric
components are regular, $\cfg$ vanishes there, but $\cffgg$ does not
unless the normal gradient of $C$ vanishes. However, since the measure
$\det{g_{MN}}$ vanishes near this axis, both $\Phi$ and $\Psi$ are
zero, and so from the point of view of solving the CR constraint
equations we need do no more here.  As discussed in the original
implementation of this method \cite{Wiseman1}, this resolves the
paradox that $C = 0$ \emph{and} $C$ has a Neumann boundary condition,
despite us solving $C$ using a boundary value formulation.  Whilst we
only impose $C = 0$ to be compatible with the elliptic equation
behaviour, if we consistently provide data on the \emph{other}
boundaries to ensure the weighted constraints vanish
\emph{everywhere}, then this Neumann condition follows automatically,
and does not need to be explicitly imposed at the symmetry axis.

Now let us consider the remaining boundaries, the horizon and
asymptotic boundary. At the horizon we only have 2 conditions from the
elliptic equations and require another to specify elliptic data.
Firstly at the horizon we find that the $\cfg$ constraint implies the
horizon temperature is a constant,
\begin{equation}
\left( A - B + C \right)_{,\chi} \mid_{\rho = \rho_0} = \frac{1}{2} \frac{J_{,\chi}}{J} \mid_{\rho = \rho_0}
\label{eq:temp const}
\end{equation}
and the remaining constraint $\cffgg$ requires,
\begin{equation}
\left( A - B + C \right)_{,\rho} \mid_{\rho = \rho_0} = \frac{1}{2} \frac{J_{,\rho}}{J} \mid_{\rho = \rho_0} - 1
\label{eq:horizbcconst}
\end{equation}

Lastly, asymptotically at large $\rho$ (or equivalently $r$) due to
the exponential decay of $\chi$ (or equivalently $z$) dependence the
$\cfg$ constraint goes exponentially to zero, and so the weighted
constraint $\Phi$ is guaranteed to go to zero, even though
$\det{g_{MN}} \sim \rho^3$. Due to this power law growth of the
measure $\det{g_{MN}}$, it is less obvious the $\cffgg$ constraint
weighted as $\Psi$ goes to zero, as $\cffgg$ (unlike $\cfg$) depends
on the homogeneous components of the metric, which only decay as a
power law. However, the behaviour of the homogeneous component implied
by the elliptic equations in \eqref{eq:asym} does also ensure that
$\Psi = 0$ asymptotically.

Now we must decide how to specify data for the elliptic equations (ie.
one condition for each metric function on each portion of boundary),
but also satisfy the constraint problem.  With the conditions already
required by the elliptic equations, we have sufficient data and $\Phi
= 0$ on all boundaries \emph{except} the horizon. The second weighted
constraint $\Psi = 0$ is satisfied only asymptotically. At the horizon
neither constraint is satisfied, and we require one more condition for
a linear combination of $A,B,C$ to make up the elliptic data as we so
far only have \eqref{eq:horizonbc}.

We could impose the constant horizon temperature condition and thus
set $\Phi$ to zero on the horizon. As we have seen $\Phi$ is zero on
all the other boundaries, and from the CR relations it obeys a Laplace
equation, so this would uniquely set it to zero everywhere. $\Psi$
would then be zero following from the CR relations and the fact it
vanishes asymptotically.

However it is numerically more stable to impose the $\cffgg$
constraint, and hence $\Psi = 0$ at the horizon instead. The $\cffgg$
constraint is a typical elliptic boundary condition, whereas imposing
the constant horizon temperature $\cfg$ constraint involves `less
local' tangential derivatives on the horizon. Now we have sufficient
data to impose both constraints globally via the CR problem. Since at
the horizon $\Psi = 0$, $\Phi$ has a Neumann condition there, and is
zero on all other boundaries, and hence will be zero everywhere as it
obeys a Laplace equation. The CR relations then imply that $\Psi$ is
constant, and must be zero as it was imposed to be zero on the horizon
and is also true asymptotically.

\subsection{How $\rho_0$ specifies the size of the black hole}

Fixing $A,B,C \rightarrow 0$ asymptotically, and thus the asymptotic
compactification radius to be $L = \pi$, there must be one constant
entering the boundary data that specifies the size, or mass of the
black hole.  Intuitively one would imagine this to be $\rho_0$ as this
certainly enters into the boundary data at the horizon, in equations
\eqref{eq:horizonbc} and \eqref{eq:horizbcconst}.

To confirm this, we must demonstrate that $\rho_0$ is a physical
quantity, and not simply a coordinate artifact. Thus we must show
there is no residual coordinate transformation (ie. conformal
transformation on $\rho, \chi$) that preserves the rectangular
boundaries, and conditions on these boundaries, and the asymptotic
radius of compactification, but changes the effective $\rho_0$ in the
new coordinates.  If this were the case, $\rho_0$ would not correspond
to a physical parameter, and therefore could not specify the size of
the black hole, which certainly is a physical parameter.

Let us suppose we have a black hole solution with a particular
$\rho_0$. Now let us construct the most general coordinate
transformation that simply preserves the rectangular boundaries. We
construct new coordinates $\rho \rightarrow \tilde{\rho} =
\tilde{\rho}(\rho,\chi)$, $\chi \rightarrow \tilde{\chi} =
\tilde{\chi}(\rho,\chi)$, and then must solve a CR problem to build
the conformal coordinate transformation. Let us do this by specifying
data for the Laplace equation that determines $\tilde{\chi}$. For the
boundaries to remain rectangular, we must specify constant
$\tilde{\chi} = \tilde{\chi}_0$ on the $\chi = 0$ boundary and
constant $\tilde{\chi} = \tilde{\chi}_{\pi/2}$ on the $\chi = \pi/2$
boundary. At the horizon, we must specify a Neumann condition on
$\tilde{\chi}$ and then the CR equations guarantee $\tilde{\rho}$ is
constant there. Regular asymptotics then give the unique solution,
\begin{equation}
\tilde{\chi} = \tilde{\chi}_0 + \left( \tilde{\chi}_{\pi/2} - \tilde{\chi}_0 \right) \frac{\chi}{\pi/2}
\end{equation}
This solution now completely determines $\tilde{\rho}$ up to a further
constant of integration.

Now that we have the general transform preserving the rectangular
boundaries, let us further restrict it by making it preserve our
boundary conditions.  Firstly, since $\tilde{\chi}_0$ does not enter
into any equation or boundary condition, we may freely set this to
zero. Secondly, since we have selected $A, B, C \rightarrow 0$
asymptotically, we require $\tilde{\chi} = \pi/2$ on the $\chi =
\pi/2$ boundary if we are not to change the compactification radius $L
= \pi$.  Thus now, $\tilde{\chi} = \chi$. This implies from the CR
relations,
\begin{equation}
\tilde{\rho} = \delta \rho_0 + \rho
\end{equation}
where $\delta \rho_0$ is a constant of integration.

However there is a subtle point. Whilst $\rho_0$ only enters the
boundary conditions explicitly at the horizon, the boundaries
conditions change on the $\chi = \pi/2$ axis for $\rho>0$ (where $C$
has a Neumann condition imposed) to $\rho<0$ (where $C = 0$ is
imposed).  Since we have fixed this transition of boundary conditions
to occur at $\rho = 0$, in the new coordinates this will occur at
$\tilde{\rho}_0 = \delta \rho_0$. Hence the transformed solution will
\emph{not} satisfy
\footnote{Note that if the physical solution had $C = 0$ and
  $C_{,\chi} = 0$ on \emph{all} of the $\chi = \pi/2$ boundary then
  $\rho_0$ would \emph{not} specify the solution. As discussed the
  constraints ensure that whilst we only impose $C = 0$, $C$ also
  satisfies a Neumann condition on the symmetry axis, $\rho<0$ and
  $\chi = \pi/2$. However, for $\rho>0$ on the periodic boundary we
  only impose a Neumann condition and there is no reason why $C$ would
  vanish there too, and indeed in the solutions we find $C$ does not
  vanish there, although it is very small.  }
the boundary conditions in the new coordinates where this transition
now occurs at $\tilde{\rho} = 0$. Thus, due to our fixing the
transition from symmetry axis to periodic boundary at $\rho = 0$,
there are no residual coordinate transformations mapping solutions
with horizon position $\rho_0$ to a transformed solution, solving the
boundary conditions but with a different horizon position
$\tilde{\rho}_0$ in the new coordinates.  Therefore $\rho_0$ does
indeed specify the physical size of the black hole.
\footnote{ It is interesting to note that the horizon must have
  spherical topology for this argument to work. If we were considering
  a string horizon, and thus chose $\rho_0 > 0$, then the boundary
  conditions would simply be Neumann all along the $\chi = \pi/2$
  boundary, and the new transformed solution in the $\tilde{\rho},
  \tilde{\chi}$ coordinates would successfully satisfy all our
  boundary conditions, but with a different value of $\tilde{\rho}_0$
  ($\ne \rho_0$) on the horizon.  It is for this rather subtle reason
  that the boundary conditions in the non-uniform string case of
  \cite{Wiseman3} must be imposed differently, using the $\cfg$
  constraint equation on the horizon (rather than $\cffgg$ as we use
  here) which, being a tangential condition, rather than a condition
  on the normal derivatives, introduces a new integration constant
  that parameterises the mass, or equivalently $\lambda$ for the
  string solution.  The fact that this is a tangential condition
  appears to make the algorithm `less local' and to require
  considerable under-relaxation, whereas here we do not need this.
  However, here while we need not damp the relaxation, the exposed
  $\rho<0$ symmetry axis does lead to the coordinate singularity
  induced stability problems discussed above.  }

\subsection{Thermodynamic quantities}
\label{sec:thermo}

We will compute the temperature $\mathcal{T}$ and entropy
$\mathcal{S}$ of the black hole solutions, and these are given as,
\begin{eqnarray}
{\cal T} &=&  \frac{1}{2\pi} \frac{3e^{A-B+C}}{2^{8/3} \sqrt{J}} \mid_{\rho = \rho_0}
\cr
\cr
{\cal S} &=&  \frac{1}{4} 
       \left[ 2 e^{4R_S} \int d\Omega_3 \int^{\pi/2}_{0} d\chi
         \sqrt{J} r^3 e^{4R+C} \right]_{\rho = \rho_0}
\label{eq:temp and entropy}
\end{eqnarray}
The mass may be computed by two independent methods.  Firstly we may
determine the mass from the asymptotics of the metric, as,
\begin{equation}
M =   \frac{\pi^2 L }{4}
          \left( - 3 a_2 - b_2 + c_2  \right),  
\label{eq:asymptotic mass}
\end{equation}
using the expansions \eqref{eq:asym}\cite{Hawking_Horowitz}.  Secondly the mass may be
determined by integration from the First Law, using $dM/d\rho_0 =
\mathcal{T}(\rho_{0}) d\mathcal{S} / d\rho_{0}$ (which applies for
fixed asymptotic compactification radius - see below) along the branch
of solutions, and taking $M = 0$ for $\rho_{0} \rightarrow - \infty$
to define the integration constant. Later (figure \ref{fig:temp_mass})
we will see very good agreement between these two values. This is a
good indication that the elliptic equations are well satisfied
globally and the boundary conditions are imposed correctly.  An
important point is that the First Law does \emph{not} test whether the
constraint equations $\crz, \crrzz$ are satisfied, and we elaborate on
this point in Appendix \ref{app:firstlaw}. Since we completely relax
the elliptic equations, it is then not terribly surprising that the
First Law holds very well. What is absolutely essential is that the
constraints are also checked, to ensure they are well satisfied. They
indeed are, and these tests are outlined in Appendix \ref{app:checks}.

Since the black hole geometries are compactified, the mass is not the
only asymptotic charge. This feature, shared more generally by branes
was originally discussed in \cite{Zamaklar}, and more recently in
relation to the black hole/black string problem in
\cite{Harmark_Obers2,Kol_Piran2,Harmark_Obers3}. It is easy to see
there must be another charge. As the geometry becomes homogeneous at
large distances from the symmetry axis it can be dimensionally reduced
to Einstein-dilaton-Maxwell theory.  For our regular static solutions
the Maxwell vector can always be gauged away, but we are left with
gravity and the dilaton scalar, indicating we should consider both an
asymptotic mass, but also a scalar charge.

From a purely 6-dimensional point of view, the second charge can be
thought of as a binding energy per unit mass, $n$, resulting in a
modified First Law,
\begin{equation}
dM = T \, dS + n \, M \, \frac{dL}{L}
\end{equation}
with the new term representing work done when varying the asymptotic
size of the extra dimension.  In our solutions, fixing $L$ we
reproduce the usual form of the First Law for black holes. However, we
may determine $n$ from the asymptotics of the metric, or from the
Smarr relation,
\begin{equation}
T S = \frac{3 - n}{4} M
\label{eq:smarr}
\end{equation}
again discussed in \cite{Zamaklar}, and given explicitly for the
problem at hand in \cite{Harmark_Obers2,Kol_Piran2},
where $n$ was calculated from the asymptotics to be,
\begin{equation}
n = \frac{a_2 + 3 \, b_2 - 3 \, c_2}{3 \, a_2 + b_2 - c_2}
\label{eq:n}
\end{equation}
As emphasised in \cite{Kol_Piran2} we may use the Smarr formula as a
check of our numerics. This is very similar to the First Law check
discussed above, which only involves the mass, whereas Smarr's law
involves both $M$ and $n$. It is important to note that for the same
reasons the First Law does not probe the constraint equations, the
Smarr formula also does not. Thus again it is no replacement for the
checking of constraint equation violations we perform in Appendix
\ref{app:checks}.

\subsection{Behaviour of the method}
\label{sec:behaviour}

\begin{figure}
\centerline{\psfig{file=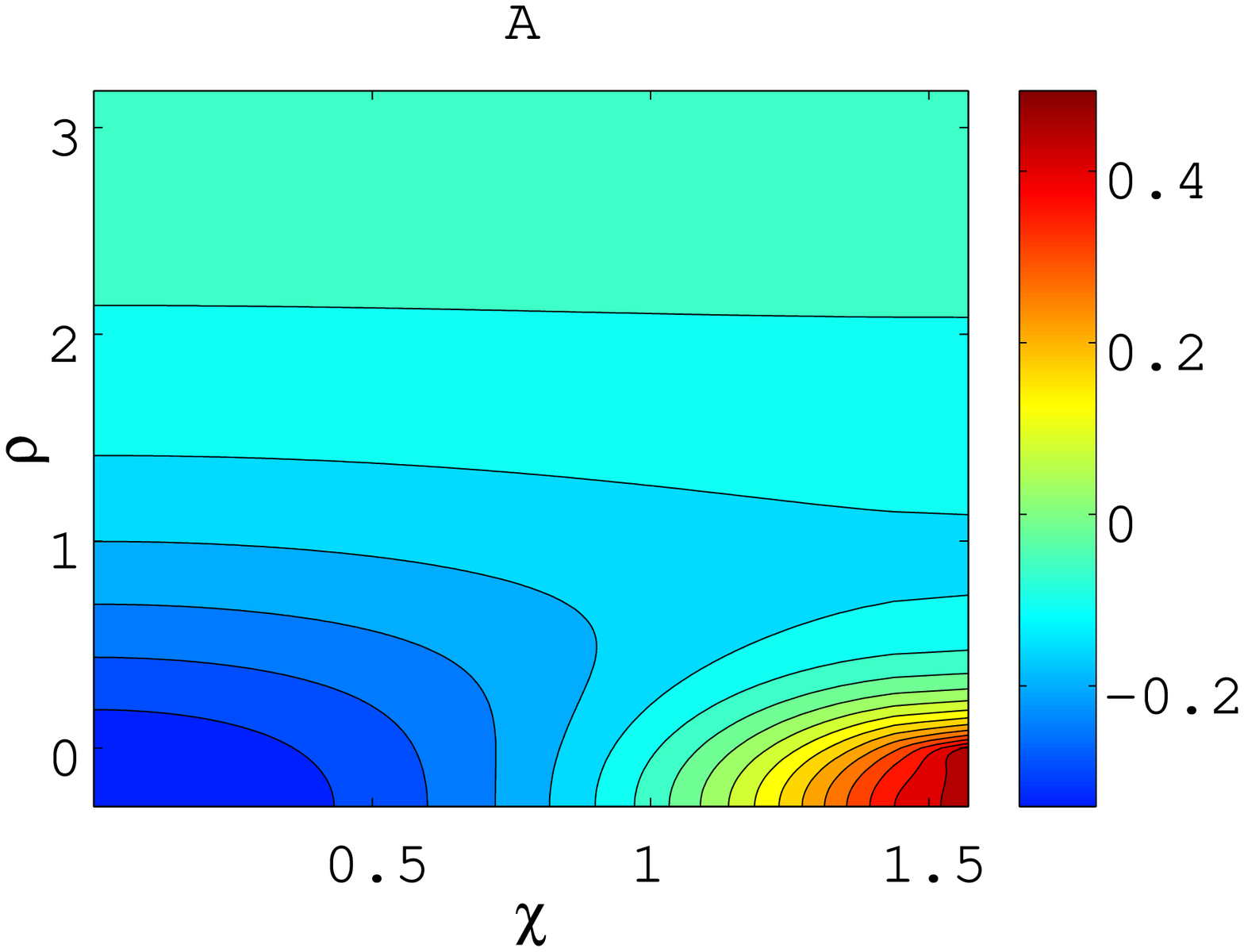,width=3.5in}\psfig{file=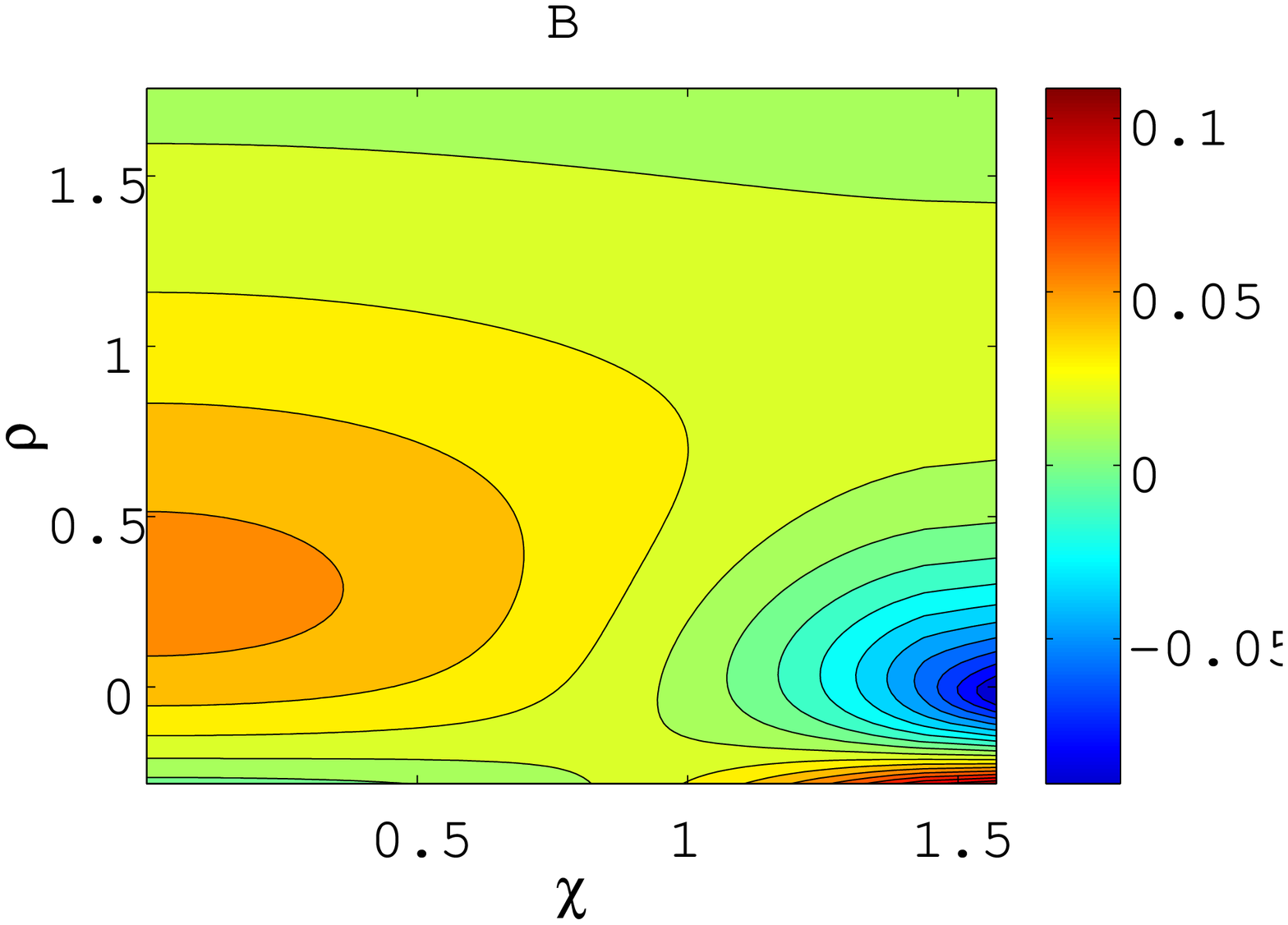,width=3.7in}}
\centerline{\psfig{file=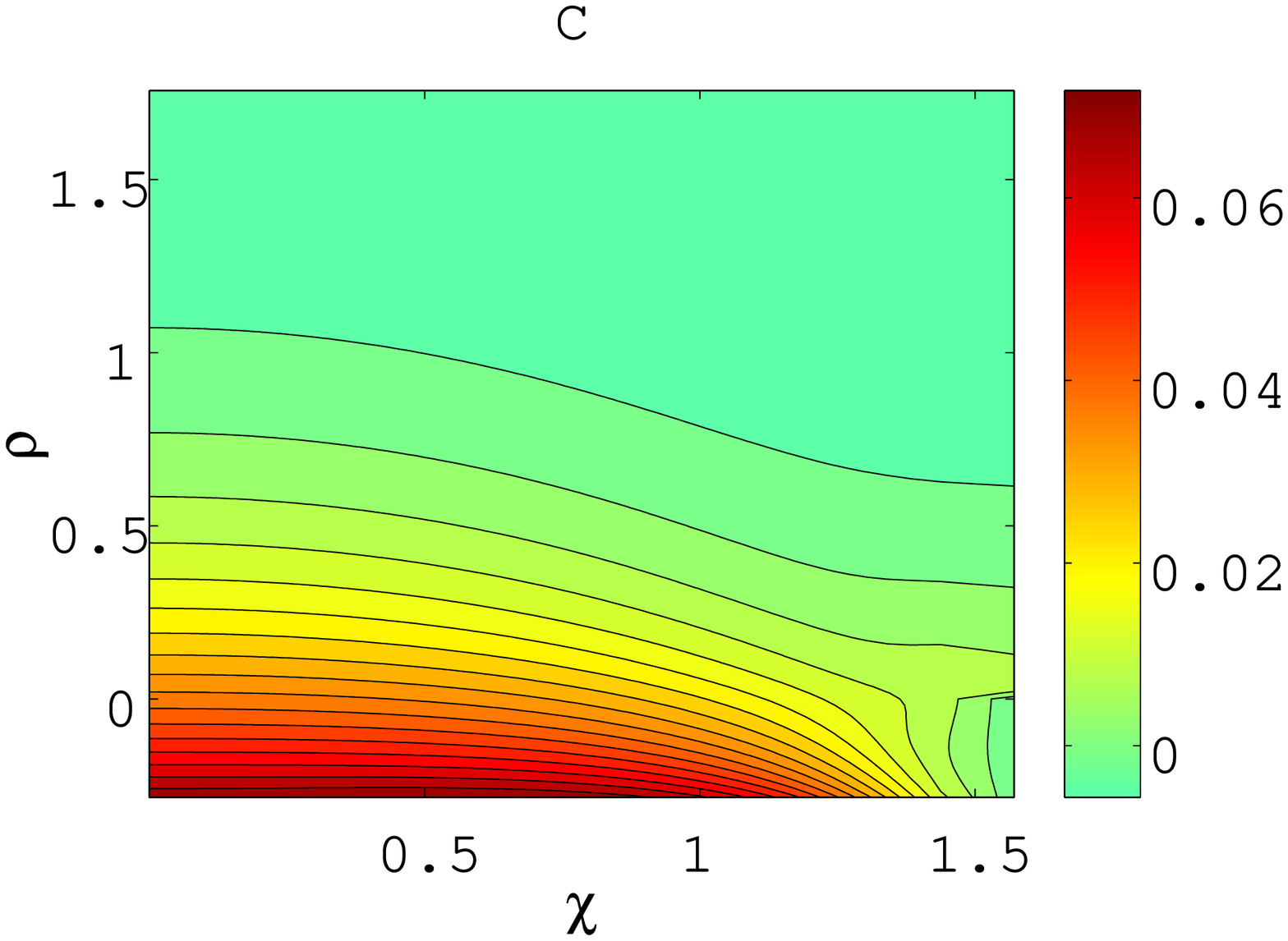,width=3.5in}}
\caption{
\label{fig:ABC}
Plots of the metric functions $A,B,C$ for a black hole solution with
$\rho_0 = -0.28$. This is quite near to the maximal size ($\rho_0 =
-0.18$) we were able to construct before we become limited by
gradients and lattice resolution near the symmetry axis at $\chi =
\pi/2$ for $\rho <0$. Already this black hole solution has equal
horizon volume and mass to the most non-uniform strings compactified
on the same asymptotic radius. Whilst the lattice is large, being
$140*420$ in $\chi,\rho$, since $\rho_0$ is very close to zero the
number of points along the symmetry axis is only around $\sim 20$.
This is still enough to see good behaviour in the metric functions.
Note that $B, C$ are much less than $A$ in magnitude.  The maximum
$\rho$ for the lattice is $\sim 5$, and not all the domain is shown as
the functions simply go smoothly to zero at large $\rho$.  }
\end{figure}

Following the above method, we may construct a unique numerical
solution for each value of $\rho_{0}$.  The elliptic equations can be
solved very stably, once the coordinate induced instability at the
axis is dealt with (see Appendix \ref{app:details}). Since we have
designed the coordinates and method to have $A, B, C \rightarrow 0$ in
the small black hole limit $\rho_0 \rightarrow - \infty$, the method
behaves well in this limit. As the black hole becomes larger, so
$\rho_0$ is finite and negative, $A, B, C$ deviate away from zero,
although remain regular as we expect.  Topology of the $\rho$
coordinate dictates $\rho_{0} = 0$ is the largest black hole that
could exist.  Using reasonable resolutions (up to $\sim$ 140*420), we
were able to find solutions with a maximum $\rho_{0} \simeq - 0.18$,
yielding a black hole horizon with typical radius comparable to the
asymptotic compactification radius.  As an example we show the metric
functions $A, B, C$ for $\rho_{0} = -0.28$ in figure \ref{fig:ABC}.
Note that the magnitude of $C$ is much less than that of $A, B$, and
this is increasingly true the smaller the black hole. Hence the
spatial sections are approximately conformally flat.

The larger the black hole is, the larger the gradients in the metric
functions, and for a fixed resolution the method no longer converges
past a certain black hole size. Going to a higher resolution we find
the problem is removed, and the size can be further increased, but
obviously the problem then re-occurs at a new larger size. The key
area where we lack resolution is near the symmetry axis.  For the
large black holes, with $\rho_0$ closer to zero, there become fewer
and fewer points there. With the maximum size we could find, $\rho_0 =
-0.18$, so the coordinate distance of the symmetry axis is $0.18$,
compared to the coordinate distance along the horizon which is $\pi/2
\sim 1.6$, and hence with our simple discretization scheme (see
Appendix \ref{app:details} for details) the axis is allocated far
fewer points.  The closer $\rho_0$ gets to zero, the more acute this
problem. Thus in our simple numerical implementation, we are limited
by resolution, and hence computation time.  We present results in this
paper using modest resources and simple relaxation algorithms.  It is
likely that with improvements in both areas one can achieve far
improved data.  For example, adaptive grid methods may circumvent the
lack of resolution near the symmetry axis, but it is a serious
challenge to implement these and maintain a stable relaxation
agorithm. However, already with our simple implementation it is
possible to derive interesting physical results as we shall see.

%
\section{Results}
\label{sec:results}
%

The questions we wish to address are whether there is an upper mass
limit for these solutions, and whether the geometry is compatible with
continuation to the non-uniform string branch.  Whilst $\rho_{0} = 0$
is the largest black hole due to the topology of the $(\rho,\chi)$
coordinate system, obviously the metric functions $A, B, C$ may
diverge in this limit, and consequently so might the horizon volume
and mass, and thus a priori we have no reason to assume such an upper
mass limit will exist. On physical grounds one could argue that a
black hole would not be able to `fit' into the compact extra
dimension, but we stress that in the unstabilised pure Kaluza-Klein
theory it is only the asymptotic radius we have fixed for the branch
of solutions, and there is nothing to prevent the geometry along the
axis and horizon from decompactifying as the black hole becomes
larger.

\subsection{Horizon geometry} 

\begin{figure}
\centerline{\psfig{file=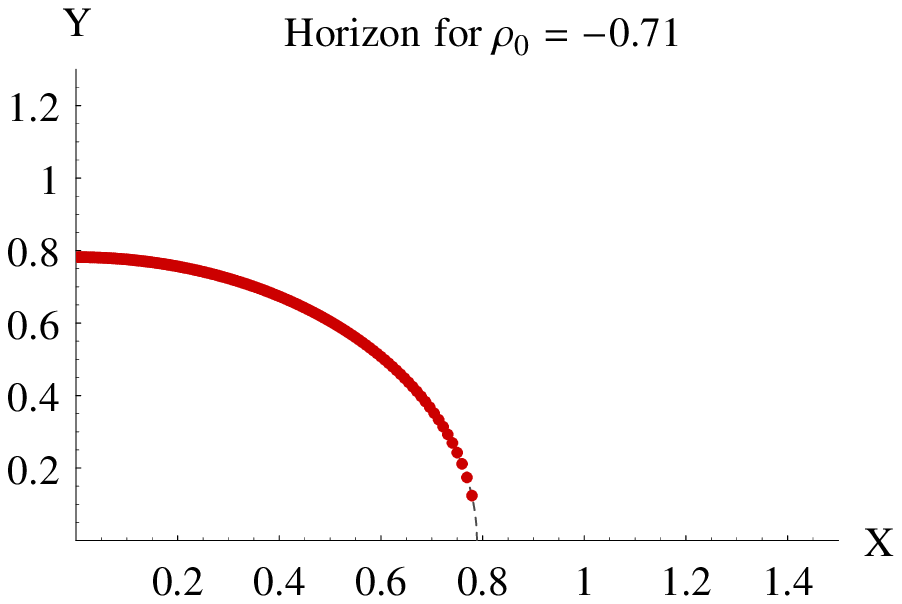,width=3.5in}
\psfig{file=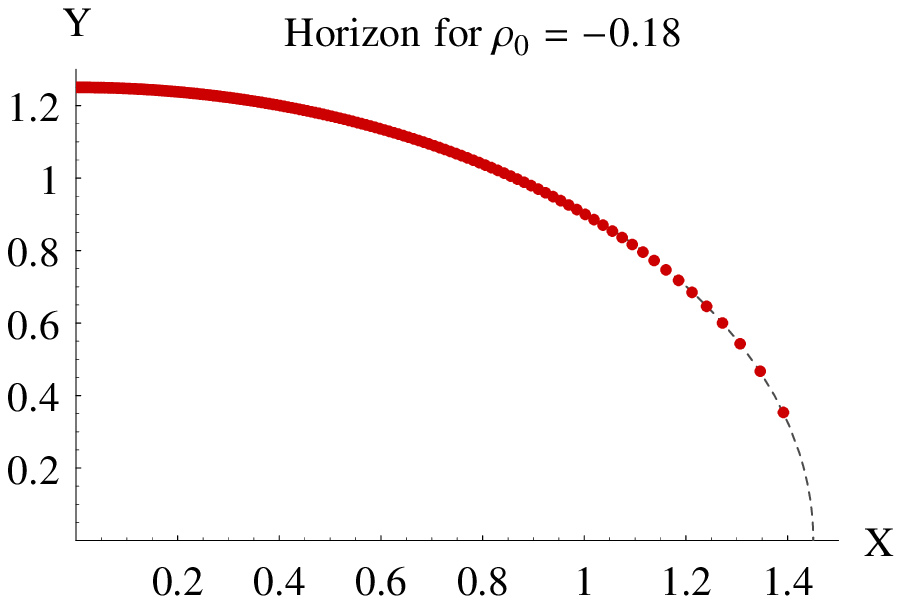,width=3.5in}}
\caption{
\label{fig:embedding1}
Embedding of horizons into Euclidean space for a moderate black hole
on the left, with $\rho_0 = -0.71$, and the maximal black hole to the
right with $\rho = -0.18$. Whilst we interpolate the numerical
functions to determine the embedding, we plot here the actual
positions of the lattice points for our highest resolution to give an
indication of how the resolution varies with position on the horizon.
For all the solutions, we may fit a prolate ellipsoid to the horizon,
the dashed line, and in all cases find perfect agreement.  }
\end{figure}

\begin{figure}
\centerline{\psfig{file=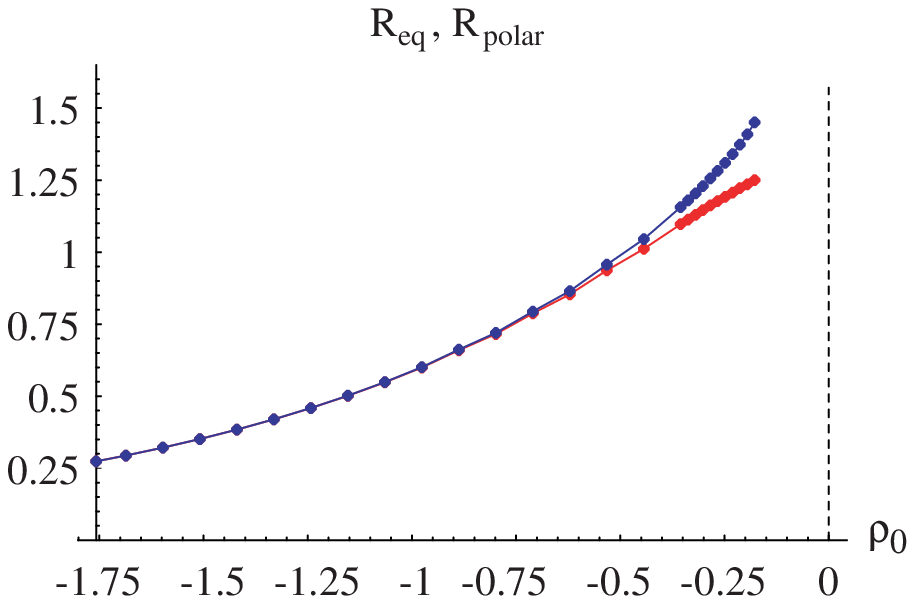,width=3.5in}\psfig{file=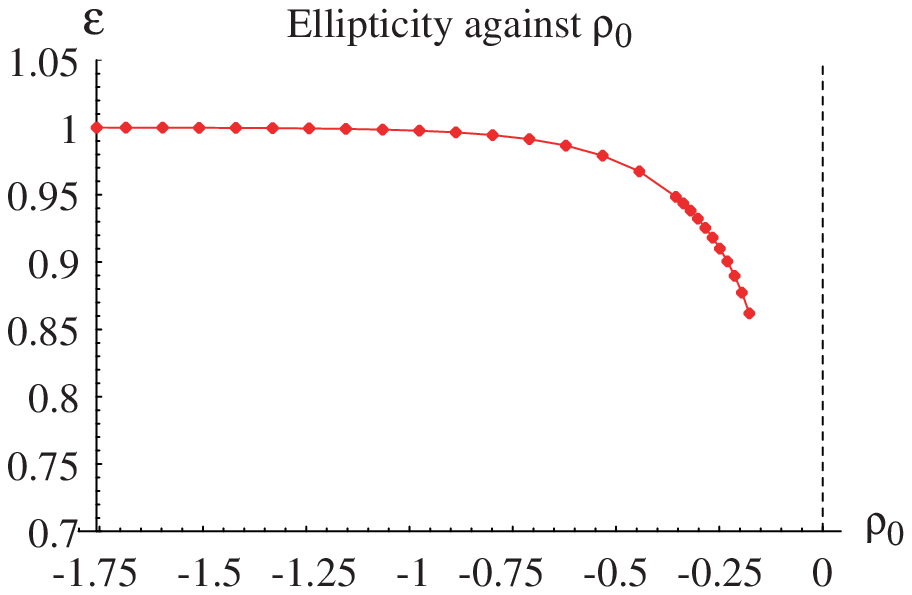,width=3.5in}}
\caption{
\label{fig:embedding2}
On the left we plot the equatorial (red) and polar (blue) radii of the
ellipses fitted to the horizon embeddings, against the parameter
$\rho_0$ specifying the size of the black hole. It is unclear what
happens in the limit $\rho_0 \rightarrow 0$ where the $\rho$
coordinate changes topology. It would be very interesting to know if
$R_{eq}, R_{polar}$ remain finite or not, and thus whether there is a
maximum mass black hole or not.  Since the geometry near the axis and
horizon decompactifies, even though the largest black holes found have
comparable radii to the corresponding asymptotic half period distance
$L/2 = \pi/2$, the horizons are still quite spherical. We plot the
ratio of the radii in the right hand diagram. Again it is unclear what
will occur in the limit $\rho_0 \rightarrow 0$.  }
\end{figure}

In order to approach these questions we embed the \emph{spatial}
horizon geometry into 5-dimensional Euclidean space,
\begin{equation}
ds^2 = dX^2 + dY^2 + Y^2 d\Omega^2_{(3)}
\label{eq:embedmetric}
\end{equation}
and matching our geometry at $\rho = \rho_{0}$ implies that,
\begin{eqnarray}
Y(\chi) & = & r \, e^{\beta + \frac{2}{3} \gamma} \mid_{\rho = \rho_{0}}
\\ \nonumber \\ \nonumber
X(\chi) & = & \int_0^\chi d\chi' \sqrt{ \left[ J(\chi') \, e^{2 ( \beta - \gamma )(\chi')} - \rho_{,\chi'}^2 \right]}_{\rho = \rho_{0}} 
\end{eqnarray}
and we interpolate the numerical data to perform the integral.
Clearly for small black holes we expect a spherical horizon, and for
larger black hole we expect deformation.  We find excellent agreement
with the horizon being a prolate ellipsoid for all our solutions up to
the maximum size available $\rho_0 = -0.18$. In figure
\ref{fig:embedding1} we plot a moderate and a large black hole to
demonstrate the accuracy of the ellipsoid fit. We plot the positions
of the actual lattice points in the embedding coordinates $X, Y$ for
our highest resolution, and against these we plot the fitting ellipse,
and note that all the points fall consistently on the fit curve. Thus
from now on it is easier for us to characterise the geometry using the
major (polar) and minor (equatorial) axis radii, which we term
$R_{polar}, R_{eq}$. Then using this elliptical fit we plot the
ellipse radii and ellipticity,
\begin{equation}
\epsilon = \frac{R_{eq}}{R_{polar}}
\end{equation}
against $\rho_0$ for all our solutions in figure \ref{fig:embedding2}.
We see that for the largest black holes the ellipticity decreases to
only $\sim 0.87$, even though the ellipsoid radius $R_{polar}$
increases to $\sim 1.5$, which is approximately the size of the
asymptotic compactification half-period $L/2 = \pi/2$. Thus it is
clear from the prolateness and lack of deformation that the geometry
around the symmetry axis is decompactifying. 

\begin{figure}
\centerline{\psfig{file=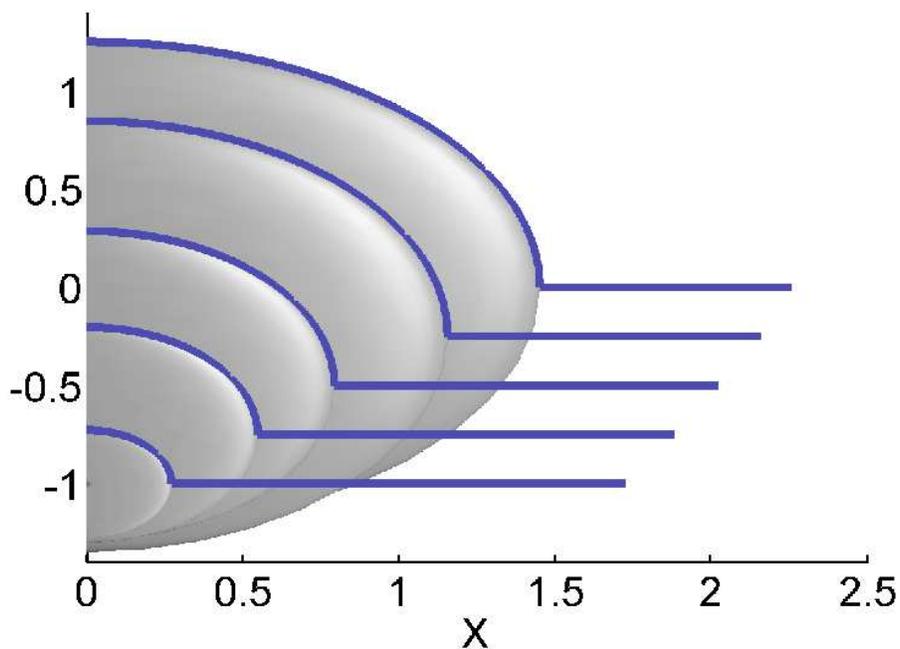,width=5in}}
\caption{
\label{fig:embedding3}
We show both the embedding of the horizon, and also the symmetry axis
for various solutions ranging up to the largest black hole found.
These embeddings are for half a period, and should be reflected about
$X = 0$ to obtain the full geometry. The asymptotic radius of this
half period is $L/2 = \pi/2$. We clearly see that the total space
taken up by the solutions in the embedding $X$ coordinate grows as the
black hole increases in size, despite the symmetry axis decreasing in
length, indicating the overall geometry near the horizon and axis
decompactifies. We also see that even the largest black hole still
appears to have considerable `room' to increase its size still
further.  }
\end{figure}

\begin{figure}
\centerline{\psfig{file=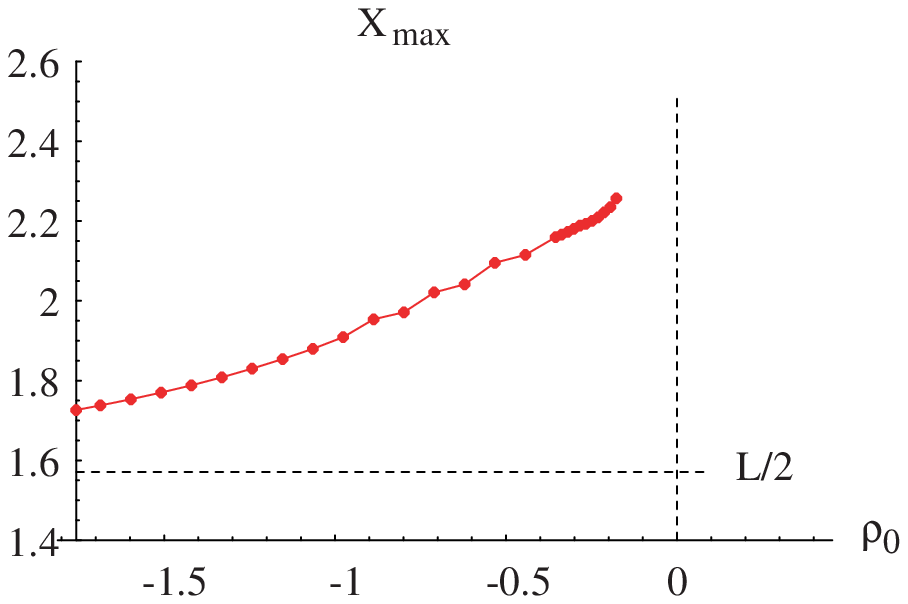,width=3.5in}}
\caption{
\label{fig:embedding4}
Plot showing $X_{max}$, the `length' taken up by half the horizon and
symmetry axis when embedding into Euclidean space. This quantity
offers a physical and coordinate invariant measure of the radius of
compactification near the horizon. We see as the black hole grows,
with increasing $\rho_0$, the axis decompactifies considerably. Again,
it is unclear what happens in the $\rho_0 \rightarrow 0$ limit. }
\end{figure}

We now wish to characterise this decompactification. In figure
\ref{fig:embedding3} we plot a selection of black hole embeddings, now
including the embedding of the symmetry axis to show its proper
length. Note that we only show half of the full period, and thus
reflecting the horizon and axis about $X = 0$ generates the full
compact period. The asymptotic compactification radius for half the
period is $L/2 = \pi/2$ here. We term the length of the axis for half
the period, $L_{axis}$, and it is given as,
\begin{equation}
L_{axis} = \int_{\rho_0}^{0} d\rho J e^{\beta-\gamma} \mid_{\chi = \pi/2} 
\end{equation}
We see $L_{axis}$ decreases in these figures for larger black holes,
but the horizon radii increase faster, resulting in an overall
decompactification. In figure \ref{fig:embedding4} we show the maximum
value of the embedding coordinate $X$ to contain half a period of both
the horizon and the symmetry axis,
\begin{eqnarray}
    X_{max} = X( \pi/2) + L_{axis}
\end{eqnarray}
which is essentially the same as $(R_{polar} + L_{axis})$ since the ellipse is
such a good fit to the horizon geometry. We take this quantity to be
the physically relevant (and coordinate invariant) measure of the
compactification length near the axis. 

These plots clearly show the axis decompactifying. Since we are only
able to `grow' black holes to $\rho_0 = -0.18$, it is unclear whether; i)
there exists a maximal mass black hole, or ii) whether the
decompactification continues in such a way that arbitrarily large
black holes may exist, eg. with $L_{axis} \rightarrow 0$ or a constant
in the limit of infinite mass. If there were a maximum mass, it seems
likely from the figures that $\rho_0$ can still be increased some way
more before we would expect to reach it.

\subsection{Comparison with non-uniform strings}

We now compare these geometries with those of the critical uniform
string ($\lambda = 0$), and the most non-uniform strings ($\lambda
\simeq 4$) found in \cite{Wiseman3}, rescaling the asymptotic radius
appropriately, and defining $\lambda$ as in that paper,
\begin{eqnarray}
    \lambda = \frac{1}{2} \left( \frac{R_{max}}{R_{min}} - 1 \right)  
\end{eqnarray}
where $R_{max,min}$ are the maximum and minimum horizon radii
respectively.  Whilst $\lambda = 3.9$ was the most non-uniform string
found there, the geometry and also thermodynamic quantities $M,
\mathcal{T}, \mathcal{S}$ appear to asymptote for large $\lambda$ (see
also \cite{Wiseman4}) and thus the $\lambda \rightarrow \infty$ values
are expected to be very similar to those at $\lambda \simeq 4$,
probably only differing by a few percent.  Strong evidence for this
comes from the realization that a conical geometry forms at large
$\lambda$, as tested in \cite{Kol_Wiseman}. Once $\lambda$ is
relatively large, say $\lambda \sim 2$, only the geometry near the
string `waist' appears to change with increasing $\lambda$ as the cone
forms.  With the rescaling so that the asymptotic radius for one
period of the solution is $L = \pi$, we find the following values,
\begin{eqnarray}
\mbox{Critical string} \qquad R_{max,\lambda = 0} & = & 0.64 
\\ \nonumber
X_{max,\lambda = 0} & = & \frac{\pi}{2}
\\ \nonumber
M_{\lambda = 0} & = & 1.47  
\\ \nonumber
{\cal T}_{\lambda = 0} & = & 0.250
\\ \nonumber
{\cal S}_{\lambda = 0} & = & 3.99
\\ \nonumber
\\ \nonumber
\mbox{Highly non-uniform string} \qquad R_{max,\lambda = 3.9} & = & 1.11
\\ \nonumber
X_{max,\lambda = 3.9} & = & 1.91
\\ \nonumber
M_{\lambda = 3.9} & = & 3.38  
\\ \nonumber
{\cal T}_{\lambda = 3.9} & = & 0.184  
\\ \nonumber
{\cal S}_{\lambda = 3.9} & = & 12.94  
\end{eqnarray}
Here $R_{max,\lambda}$ is the maximal radius of the horizon, and
$X_{max,\lambda}$ is again the maximum value of $X$ when embedding
half a period of the strings into Euclidean space, using the metric
\eqref{eq:embedmetric} as for the black holes, and taking $X = 0$ at
the maximal radius of the horizon, so $X = X_{max}$ at the `waist',
the minimum radius. Now, there is no exposed symmetry axis, so
$X_{max}$ is just the change in $X$ when traversing the horizon for a
half period.  As noted in \cite{Wiseman4} the proper distance (which
is not equal to the embedding coordinate $X$) along the horizon
increases with $\lambda$ indicating the geometry decompactifies there.

\begin{figure}
\centerline{\psfig{file=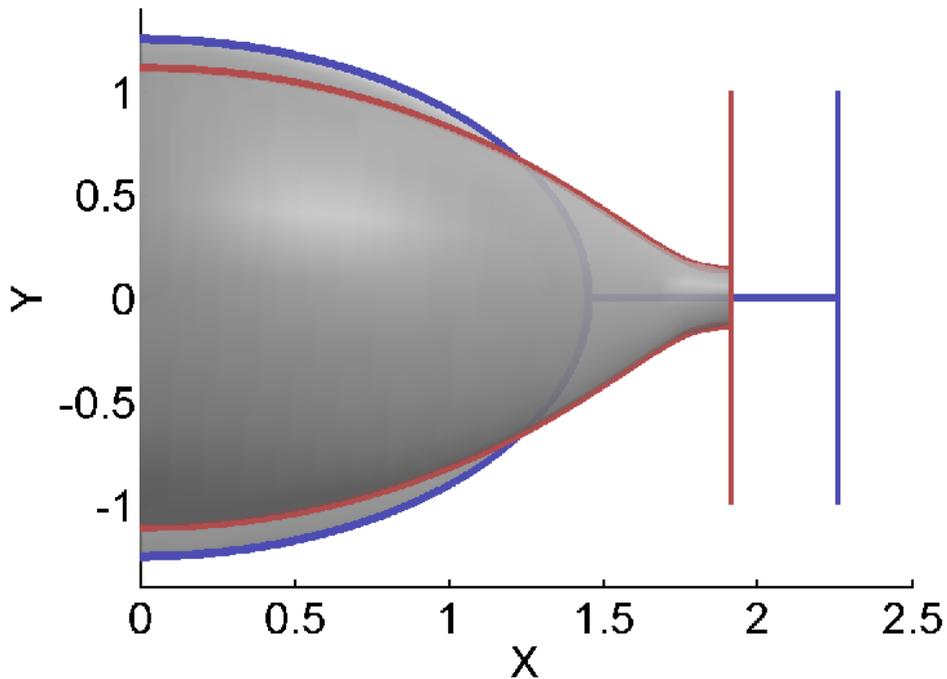,width=5in}}
\caption{
\label{fig:string_bh}
An illustration of the intrinsic geometries of the maximum sized black
hole found ($\rho_0 = -0.18$) compared to a highly non-uniform string
($\lambda = 3.9$) whose geometry is expected to be very close to the
limiting $\lambda = \infty$ solution, with a conical geometry near the
waist of the horizon. Only half a period of the solutions is shown,
and for the full geometry one should reflect about $X = 0$. Both are
taken to have an asymptotic compactification half period $L/2 =
\pi/2$. We see immediately that the black hole is simply `bigger' in
all aspects. It has larger equatorial radius, and including the
exposed symmetry axis, it takes up more `room' in the $X$ direction -
ie. it has decompactified more. Indeed we also find it has a higher
mass, and lower temperature. Furthermore, it looks very much as if we
can increase the black hole size further still. The implication is
that the $\lambda = \infty$ string solution presumably cannot be
connected (through a conical topology change) to \emph{this} branch of
black hole solutions. }
\end{figure}

The key observation of this paper is now evident from the previous
plots of the embedded black hole geometry. Already for $\rho_{0}
\simeq -0.35$ we see the black hole equatorial radius $R_{eq}$ is
equal to $R_{max}$ for the highly non-uniform $\lambda = 3.9$
non-uniform string, which as stated above we take to be approximately
equal to $R_{max}$ in the limit $\lambda = \infty$.  And as $\rho_{0}$
increases past this point, the black holes continue to become larger
in radius.  In addition we also see from $X_{max}$ that the geometry
along the axis has decompactified as much as that of the very
non-uniform strings for even quite small black holes with $\rho_0
\simeq -1.0$, and again the trend seems to continue for increasing
$\rho_{0}$ past this point.  The implication is clear, that it seems
difficult to imagine the black hole solutions making a transition at
$\rho_0 = 0$ via a cone geometry to the most non-uniform strings, when
they simply become `bigger' than the most non-uniform strings already
for $\rho_{0}\simeq-0.35$.  We may gain more insight into this result
in figure \ref{fig:string_bh} by plotting the embedding of the
$\lambda = 3.9$ non-uniform string, and the largest black hole relaxed
($\rho_0 = -0.18$) into Euclidean space, including the symmetry axis
of the black hole. Again we note that with increasing $\lambda$ past
3.9 we only expect the geometry in the cone region to change, and thus
the geometry of the $\lambda = 3.9$ string should be very close to
that of the limiting string at $\lambda = \infty$ (see \cite{Wiseman4}
for curves of $R_{max}$ against $\lambda$).

Whilst we earlier claimed that resolution becomes limited near the
horizon and axis for the large black holes, we find that the values of
$R_{eq},R_{polar}$ only vary by $\sim 1 \%$ when doubling the
resolution from $70*210$ to our highest resolution $140*420$ for the
solution with $\rho_{0} \simeq -0.35$ that parallels the most
non-uniform string horizon size. The length along the axis $L_{axis}$
varies only a little more, around $\sim 3 \%$. Thus while decreasing
axis resolution with increasing black hole size limits the ability of
the algorithm to converge, the resolution is still high enough for the
accuracy of our large black hole solutions to be high. For further
comparison of quantities measured at different resolutions, see figure
\ref{fig:temp_mass} and the Appendix \ref{app:details}.

\subsection{Thermodynamics}

\begin{figure}
\centerline{
    \psfig{file=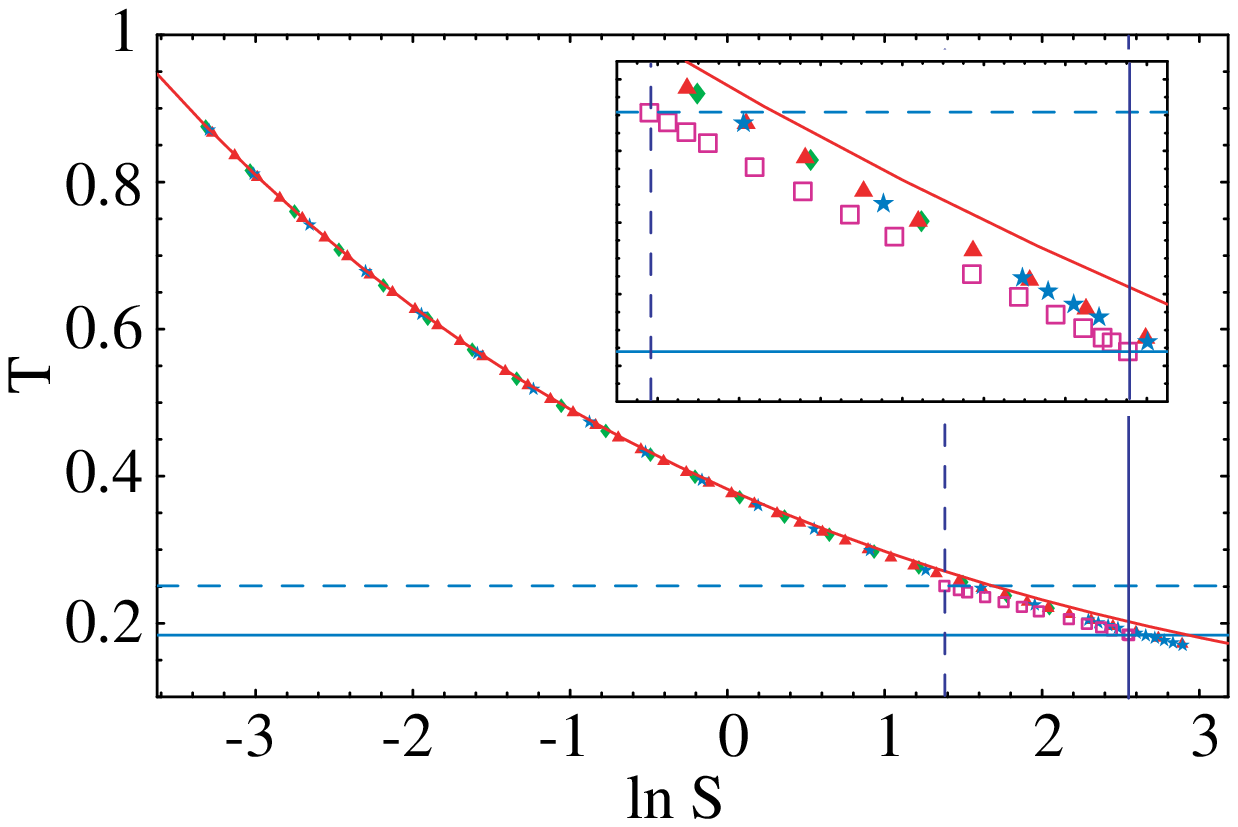,width=5in}}
\centerline{    \psfig{file=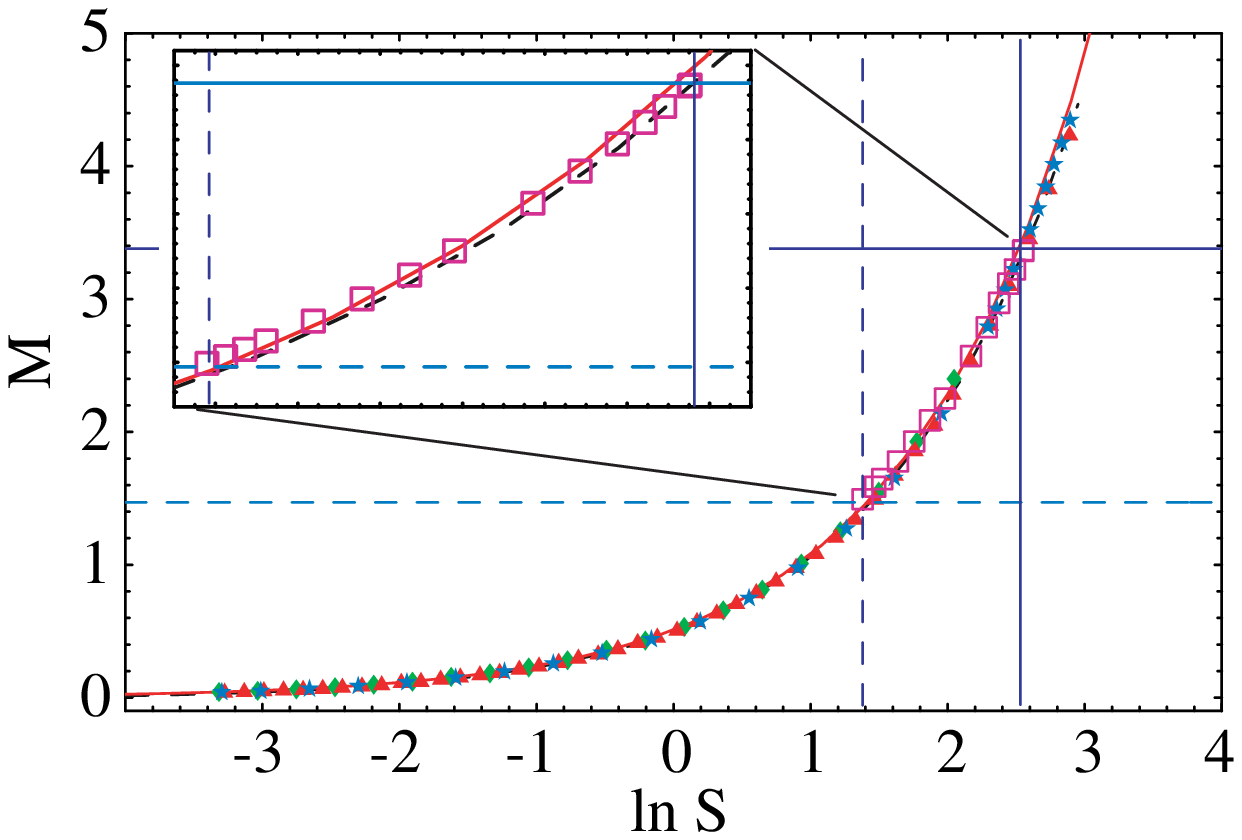,width=5in}
}
\caption{
\label{fig:temp_mass}
Plot showing black hole temperature $\mathcal{T}$ and mass $M$ against
entropy $\mathcal{S}$ calculated by eqns. (\ref{eq:temp and entropy})
and (\ref{eq:asymptotic mass}).  Three resolutions for the black holes
are used, $36*106$ in green, $70*210$ in red, $140*420$ in blue, and
excellent consistency is found. In the figure for the mass, the dashed
black line line gives the mass computed by integration from the First
Law, showing excellent agreement with the asymptotically measured
mass. The solid red lines give the behaviour of a 6-dimensional
Schwarzschild solution, and for these black holes, despite their size
becoming equivalent to the compactification radius, we see little
deviation from this.
We also show the same quantities plotted for the non-uniform strings
as pink open squares. The dashed and solid straight lines show the
limiting values for $\lambda = 0$, and $\lambda \rightarrow \infty$.
We include magnifications of the region where the non-uniform strings
exist. Interestingly the non-uniform values lie very close to the
black hole curves, although this appears simply to be coincidence.  }
\end{figure}

Now we turn to the thermodynamic quantities to see whether our
comparison of the black hole/string horizon intrinsic geometry is
parallelled in these independent observables. We might now reasonably
expect the mass of the black holes to become greater than that of the
most non-uniform strings, and the horizon temperature to become less.
In figure \ref{fig:temp_mass} we plot the temperature, $\mathcal{T}$,
and mass $M$ of the black hole solutions, now against the horizon
entropy $\mathcal{S}$. The mass is computed in two ways, firstly
asymptotically from the metric [see equation \eqref{eq:asymptotic mass}],
and secondly by integration from the First Law. We clearly see very
good agreement for these as expected.  As discussed earlier in section
\ref{sec:thermo}, this is a good test of the elliptic equations, but
does not test the constraints which are the really important
quantities to check for this elliptic method, as they are not imposed
directly.  These constraints are tested explicitly in Appendix
\ref{app:checks}.

On these plots we also show the same quantities for the non-uniform
string branch up to $\lambda = 3.9$. Again we emphasise that the
$\lambda = 3.9$ point probably lies very close to the $\lambda =
\infty$ point on this diagram. Our expectations are confirmed, with
the temperature becoming lower, and the mass becoming higher than the
most non-uniform strings already by $\rho_0 \simeq -0.30$, and the
trend continuing for larger $\rho_0$, again reflecting the fact that
the black holes become `bigger' than the non-uniform strings. We also
plot the behaviour of a 6-dimensional Schwarzschild solution, and find
that since the axis is decompactifying, and as a consequence the black
hole horizon geometry is only slowly deforming from being a sphere,
the black holes closely reproduce this Schwarzschild behaviour.

There is one further point we may observe from these plots.  The
non-uniform branch presumably terminates at $\lambda = \infty$, very
close to the $\lambda = 3.9$ solutions plotted, so the whole
non-uniform branch of solutions $\lambda = 0$ to $\infty$ then appears
to lie very close to the black hole curves both in $\mathcal{M}$ and
$\mathcal{T}$ against $\mathcal{S}$. Since the branches do not appear
to connect, we can think of no particular reason why this should be
so, and presumably it is just an interesting coincidence.

\begin{figure}
\centerline{\psfig{file=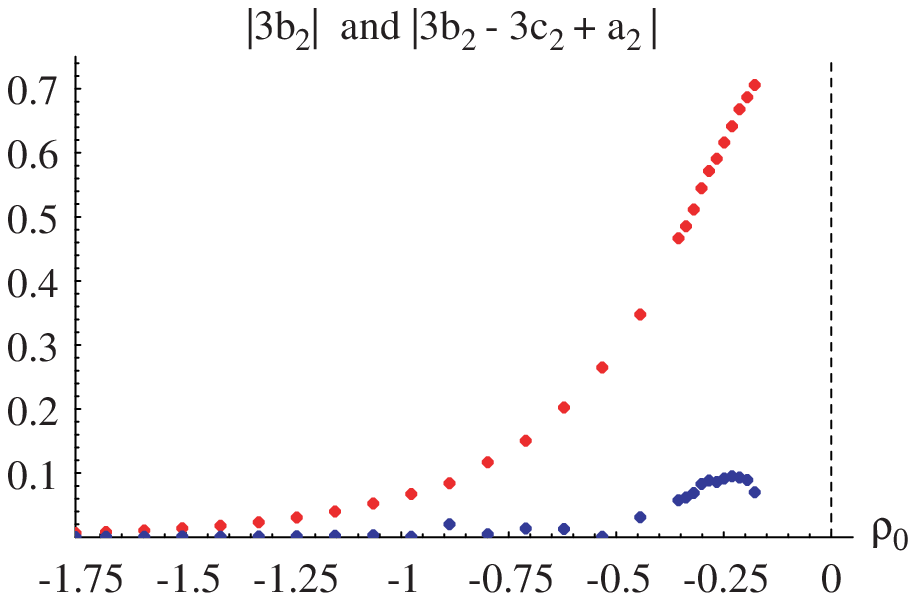,width=3.5in}\psfig{file=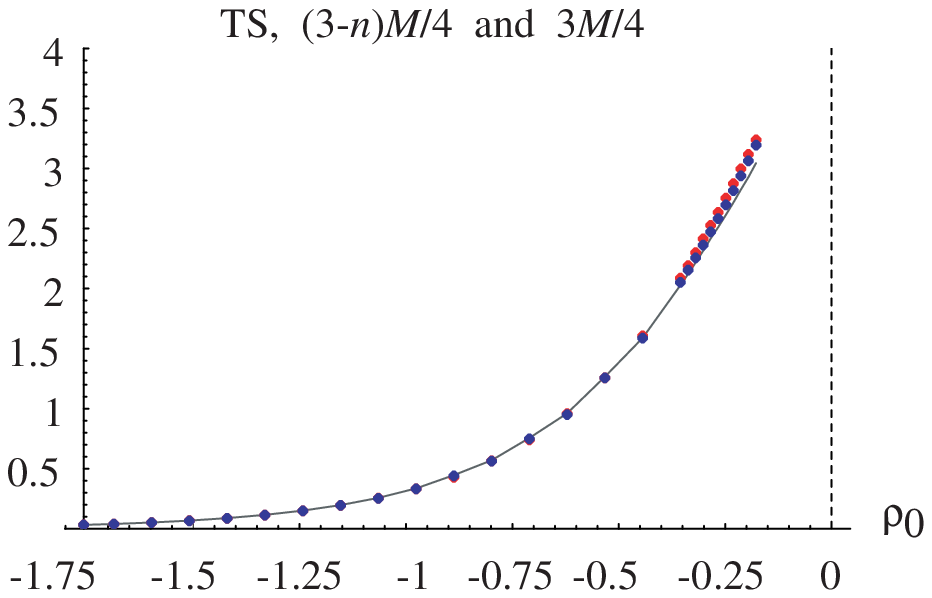,width=3.5in}}
\caption{
  Left; A plot of the magnitude of a term in the numerator of `n', the
  binding energy, and the magnitude of the numerator itself. We see
  this numerator is $\sim 10\%$ of the value of one of its constituent
  terms, and thus detailed cancellations occur, and the value of `n' is
  very small. Since the asymptotic quantities $a_2, b_2, c_2$ are
  already difficult to compute accurately, we should not trust a
  numerical quantity that depends on detailed differences of these,
  and thus is very small in comparison.  Right; A plot of the left-
  ($T S$ as a solid black line) and right-hand ($(3-n) M/4)$ as blue
  points) sides of Smarr's formula, confirming that whilst $n$ is very
  hard to determine accurately, Smarr's formula is very well
  satisfied by our data. To confirm that `n' is indeed extremely
  small, we show the right-hand side computed with $n = 0$, ie. $3
  M/4$ as red points, which gives a slightly worse fit, but only
  marginally.
\label{fig:smarr}
}
\end{figure}

Finally we turn to the last (and truly higher dimensional)
thermodynamic quantity, the binding energy $n$ of our solutions. This
turns out to be very small, and thus prone to numerical error.
Referring back to \eqref{eq:n} we recall $n$ to be given by a ratio of
asymptotic quantities, with numerator $3 \, b_2 - 3 \, c_2 + a_2$. In
figure \ref{fig:smarr} we plot the magnitudes of both a term in the
numerator, and the numerator itself, and demonstrate that the
numerator is relatively very small, $\sim 10 \%$ of the terms making
it up, indicating cancellations occur between the terms.  This is a
problem numerically since the asymptotic quantities $a_2, b_2, c_2$
are already difficult to measure, and thus a quantities depending on
detailed cancellations between them certainly should not be trusted. A
further caveat is that for small black holes both the numerator and
denominator are small, and their ratio is consequently extremely
unreliable. Computing $n$ from the asymptotics we find its value to be
less than $\sim 0.1$ for these solutions, but the errors appear large,
and we stress simply that it is small, and we do not feel we can give
its value with certainty here. We may reassure ourselves that whilst
$n$ is `noisy' in the numerical error, this is simply because $n$ is
close to vanishing, and Smarr's law is extremely well satisfied. Also
in figure \ref{fig:smarr} we plot the two sides of the Smarr relation
\eqref{eq:smarr}, $T S$ and $(3-n)M/4$, against $\rho_0$. In addition,
we also plot $3 M/4$, ie.  setting $n = 0$, on the same plot, which
lies so close to the curve with the actual measured $n$, that it is
clear $n$ is quantitatively small, and furthermore we cannot currently
expect to measure it with any accuracy as argued above. Much higher
precision would be required for this, and thus we leave determining
the actual (small) values of $n$ for future work.

%
\section{Discussion and Outlook}
\label{sec:discussion}
%

We have shown that static black holes in pure 6-dimensional gravity
compactified on a circle, ie. Kaluza-Klein black holes, may be found
using elliptic numerical methods. As expected, for fixed asymptotic
compactification radius, the small black holes behave as 6-dimensional
Schwarzschild solutions. As they grow the geometry on the axis
decompactifies relative to the fixed asymptotic radius, and the
horizon deforms to a prolate ellipsoid. Since we are limited by
numerical resolution, we are currently unable to probe whether the
axis decompactifies indefinitely, and consequently black holes of any
mass can be found, or whether instead there is an upper mass limit for
the black holes.  This is clearly an interesting issue to resolve in
future work.

The most interesting result we find is that whilst we are only able to
compute black holes whose radii become approximately equal to the
asymptotic compactification radius, these are already sufficiently
large that they are simply `bigger' than the most non-uniform string
solutions constructed in \cite{Wiseman3}, both in terms of horizon
volume and mass. For the largest black hole solutions we found the
horizons are still quite spherical since the axis geometry
decompactifies, making `room' for the horizon. It therefore appears
that they should be able to further increase in size and mass, past
the point where we are currently able to construct them, before any
possible upper mass limit would be reached. Given these results, it
therefore seems rather unlikely that this black hole branch of
solutions can merge with the non-uniform strings via a conical
geometry developing at the polar regions of the horizon, as suggested
by Kol \cite{Kol1}. This is despite the fact that we have excellent
numerical evidence that the highly non-uniform strings do indeed
exhibit the required conical geometry at their waist, which previously
lent weight to this conjecture \cite{Kol_Wiseman}.

This is then very interesting geometrically, and raises the obvious
question of whether the static non-uniform string solutions can be
continued through the $\lambda \rightarrow \infty$ solution with its
conical waist to a \emph{new} branch of black hole solutions.
Obviously while these would have horizons that do not wrap the extra
coordinate, they would be distinct solutions from the ones we construct
here, and at low mass (if they have a low mass limit) would presumably
not look like the 6-dimensional Schwarzschild solution.
\footnote{For example the non-uniform string branch may join a branch
  of black holes that then undergoes additional topology changing, and
  so in the limit of small mass the horizons are Schwarzschild-like,
  but with more than one horizon per compactification period.  }
Similarly, if the black hole branch we partially construct here does
turn out to have an upper mass limit, can this branch of solutions be
continued through to a new string solution, distinct from that
connected to the Gregory-Laflamme critical uniform string? We refrain
from speculating on these questions (see the recent
\cite{Harmark_Obers3} for an interesting discussion of a variety of
possibilities), deferring these issues until improved numerics can be
performed that confirm the current results, and can extend the range
of these elliptic methods so these questions may be tackled directly.
We do make one further general comment here. As we have seen comparing
the current work here with the previous work constructing the
non-uniform string solutions \cite{Wiseman3}, the fact that the axis
of symmetry is exposed for the black hole solutions completely changes
the boundary conditions imposed on the problem. Thus, without proof,
it would be dangerous to assume that continuation of a branch of
solutions through topology change at a conical region must always be
possible.

Whilst our computations were performed in 6-dimensions, in order to
make contact with previous work constructing the non-uniform strings,
it would be good to check the same behaviour occurs in 5-dimensions.
Whilst the difference between 4-dimensions and more than 4 is very
large, due to the additional curvature terms entering the Einstein
equations from the rotation group of the axial symmetry, the
difference between 5 and 6 dimensions is simply in coefficients
entering these equations.  Hence we would be very surprised if the 5
and 6-dimensional systems behaved qualitatively differently, but to be
sure it would be good to check by constructing the non-uniform string
and black hole solutions in 5-dimensions and comparing them.

Our findings appear to be strongly related to the decompactification
of the geometry near the horizon and axis. The reason the black holes
become larger than the non-uniform strings is because the axis
decompactifies making room for them. Therefore in order to make
contact with realistic phenomenology, and thus really determine
whether there are interesting strong gravity effects of
compactification, it is clearly important to consider the problem
again, but include some radius stabilising mechanism. Presumably once
stabilisation is included, an upper mass limit for the black holes
should be inevitable as the axis can't decompactify so easily. This is
a sufficiently important phenomenological question that this should be
checked explicitly, rather than just assumed, as if it turned out not
to be true, or only be true for certain stabilisation mechanisms, this
might provide new physical and observational constraints on
compactifications that are totally independent of the familiar weak
field constraints. It may also provide an important testing ground for
the non-linear dynamics of stabilisation mechanisms. Additional
matter, such as is required for stabilisation, simply adds elliptic
equations to the problem, and no further constraints, and thus in
principle can be easily incorporated. We note that, at least with weak
stabilisation, the Gregory-Laflamme instability will occur as usual.
Hence non-uniform string solutions will also exist, although of course
it is then not obvious that they can be deformed to have a conical
region of their horizon as in the unstabilised case. If they do behave
in the same manner as for the unstabilised theory, and if, since the
axis could not decompactify readily, the black holes are forced to
have an upper mass limit at a lower mass than those probed in this
paper, then possibly the topology change to the black hole branch that
Kol suggests could occur after all.

%
\section*{Acknowledgements}
%

We would like to thank Troels Harmark, Takashi Nakamura, Jorge Pullin,
Harvey Reall and Takahiro Tanaka, and in particular Barak Kol and
Evgeny Sorkin for interesting and valuable discussions.  Also
discussions during and following the YITP workshop YITP-W-02-19 were
useful.  TW would like to thank YITP, Kyoto and in particular Takahiro
Tanaka for much hospitality whilst some of this work was completed.
HK is supported by the JSPS and a Grant-in-Aid for the 21st Century COE, and TW was supported by Pembroke college,
Cambridge for most of the duration of this project. Numerical
computations were carried out at the Yukawa Institute Computer
Facility.

\appendix

%
\section{Appendix: Numerical details}
\label{app:details}
%

As in the previous works \cite{Wiseman1, Wiseman3, Kudoh1} we use a
simple Gauss-Seidel method with second order differencing to relax the
elliptic equations. The non-linear source terms are fixed, the
resulting Poisson equations are relaxed, and then the sources are
updated with the new solutions, and the boundary conditions are
refreshed. Repeating this, the elliptic equations are either
\emph{completely} relaxed and we find a solution, or convergence is
lost at some point early in the relaxation, and all the metric
functions diverge dramatically to nonsense. We impose the asymptotic
boundary at finite $\rho = \rho_{max}$ which we typically take $\sim
5$, so several multiples of the half periodic compactification radius
$L/2 = \pi/2$. In the next Appendix we show data for varying
$\rho_{max}$ and demonstrate this has been taken large enough so as to
be irrelevant.

Essentially all our numerical problems come from the symmetry axis $r
= 0$. Firstly, rather than discretising the grid in the $\rho,\chi$
coordinates, generically one gains stability using $\rho, \xi$ with
\begin{eqnarray}
\xi = 4(\chi - \pi/2 )^2, 
\label{eq:define xi}
\end{eqnarray}
since at $\chi = \pi/2$, all fields are even in $(\chi - \pi/2)$ and
therefore linear in $\xi$. This was used successfully in the black
hole on an RS brane to improve stability \cite{Kudoh1}.  However, even
with this modification the algorithm is horribly unstable, and even
for the smallest black holes we find no convergence. The same problem
was encountered in the earliest application of this method
\cite{Wiseman1}.  When dealing with a spherically symmetric scalar
field $\phi$ in polar coordinates, one is very familiar with terms
such as $\frac{1}{r} \partial_r \phi$ in the equations of motion.
Whilst for an elliptic relaxation these look as if they might destroy
convergence, in reality they are not a severe problem, as long as the
Neumann condition on $\phi$ is imposed at $r =0$. However, the ansatz
\eqref{eq:simplemetric} or \eqref{eq:jacmetric} generates more
singular terms in the field equations. The exact form of our metric
ansatz \eqref{eq:jacmetric} guarantees that only one of the elliptic
equations is effected, but we find that in the equation for $C$,
\begin{equation}
 J\Delta C  =   \frac{3J}{r^2} \left( e^{-10 C/3} - 1 \right) + \ldots
\end{equation}
where $J\Delta = J\left( \partial_r^2 + \partial_z^2 \right)
=(\partial_\rho^2 + \partial_\chi^2 )$.  The remaining terms have the
more usual $1/r$ multiplying derivatives.  Obviously the above term is
finite as $C \sim r^2$ near the axis.  However since we do not impose
that $C$ goes quadratically near the axis, and instead it emerges from
a combination of the elliptic equations and the constraints, during
the early stages of the relaxation this term generically destroys
convergence.

We deal with this term as in \cite{Wiseman1}. The second derivative
terms in the constraint equation $\cfg$ are simply $C_{,\rho\chi}$ and
thus have characteristics compatible with integrating $C$ over the
$(\rho,\chi)$ domain.  However this equation has no such singular term
as that above, and any solution for $C$ integrated away from the $r
=0$ axis has very good quadratic behaviour in $r$ near there.
\footnote{One might imagine using just $\cfg$ and 2 elliptic equations
  for $A, B$, but we have been unable to find a scheme like this that
  worked in practice due to the non-local nature of the integration
  for $C$.}
Thus using this constraint, we integrate for $C$, but call this
function $C2$. Since it has very good properties near the $r$ axis, we
calculate the one singular term in the elliptic equation for $C$ using
this function $C2$, rather than $C$. Whilst this seems circular in
nature, and we offer no proof why this should converge so that $C =
C2$ finally, in practice this does indeed happen, and the method
becomes very stable. Given the characteristics we need two initial
data surfaces, one at constant $\chi$, the other at constant $\rho$.
For $\chi$ we fix $C2 = C$ at $\chi =\pi/2$, which includes the $r =
0$ axis for $\rho<0$, and this ensures the quality of the $C2$
behaviour is good near the symmetry axis. For the other initial
surface we take the horizon, and fix $C2$ using the condition
(\ref{eq:temp const}) (but now for $C2$) that the horizon temperature
is a constant, giving
\begin{eqnarray}
 C2(\rho_0, \chi) = (B-A) - (B-A)|_{\chi= \pi/2} + \frac{1}{2} \log \left( \frac{J}{J|_{\chi=\pi/2}}\right) \,.
\end{eqnarray}
As discussed earlier, this condition is not used when solving the
elliptic equations, where we instead use the $\cffgg$ constraint.
However, it gives a more stable initial boundary condition for the
$C2$ integration than simply setting $C2 = C$ there directly (which
destroys convergence). We then integrate $C2$ from these two
boundaries by quadrature,
\begin{eqnarray}
C2 = \int^{\rho}_{\rho_0} d\rho \int^{\chi}_{\pi/2} d\chi~F(\rho,\chi) 
    + C(\rho_0, \chi) + C(\rho, \pi/2)\,,
\end{eqnarray}
where $F$ is the `source' term in the constraint equation.

The reason this method to compute the singular term in the elliptic
equation for $C$ appears to work is that in practice the contribution
$(e^{-10 C/3} - 1)/r^2$ is only significant near the $r = 0$ axis, and
away from there this source term dies away more quickly than the other
terms as it is suppressed by $1/r^2$. Thus whilst the process appears
very non-local, involving integration over the lattice during the
relaxation, which is not very `gentle' and might destroy convergence,
actually it only has an effect localised at the axis. Furthermore the
function $C$ is generically much smaller than the other metric
functions $A, B$, so $C$ appears to have relatively little effect on
the solution anyway.

\begin{figure}
  \centerline{\psfig{file=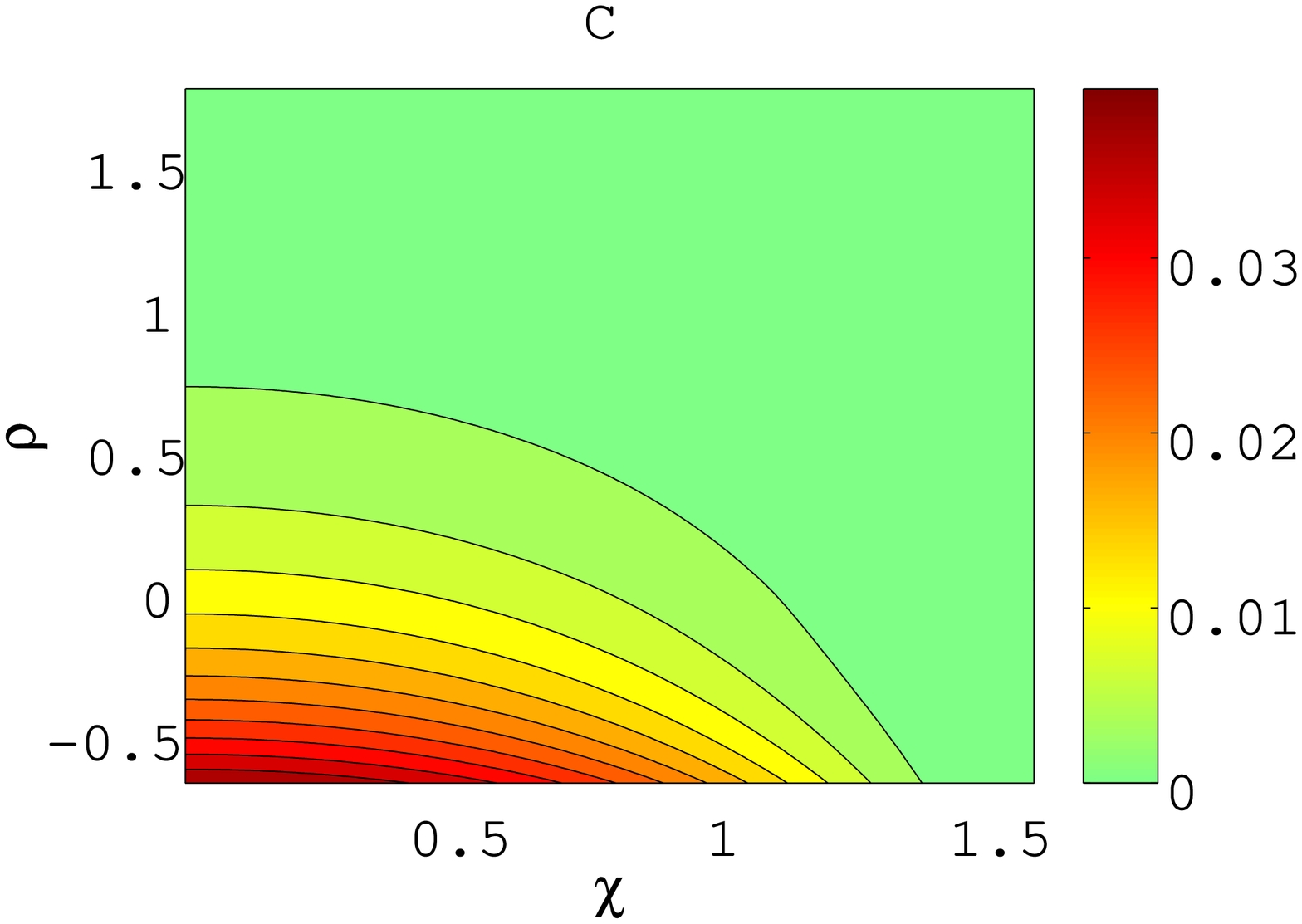,width=3.5in}
    \psfig{file=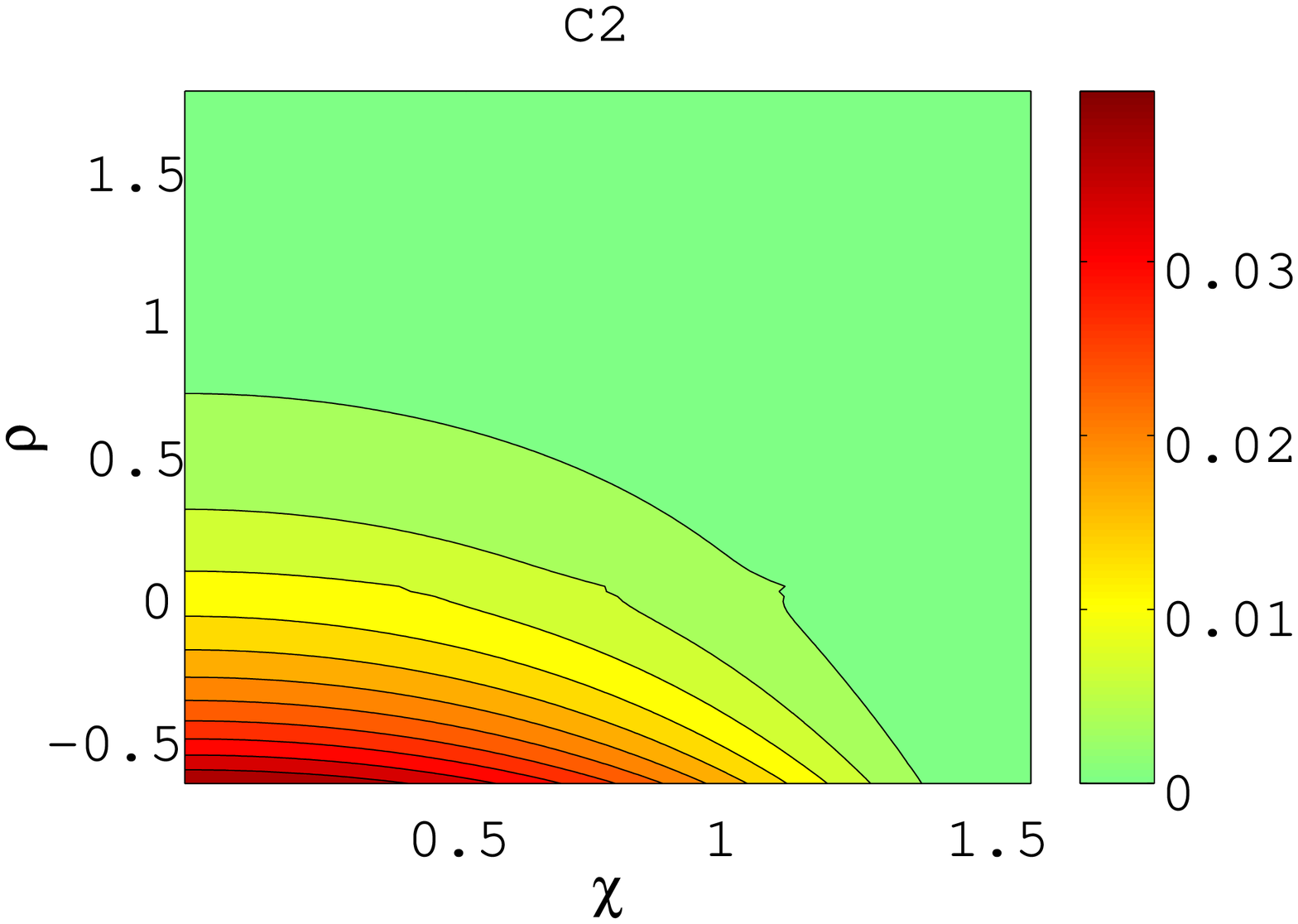,width=3.5in}}
\caption{
\label{fig:C2C}
Plots of the metric functions $C$ and $C2$ for $\rho_0=-0.71$.
Main error of $C2$ originates from the coordinate singularity at $\rho=0, \chi=\pi/2$, but its effect for $C$ through the elliptic equation is small, and the difference of their magnitude is very small compared to $C$.
}
\end{figure}

Since we do find stable converged solutions we may check that $C$ is
equal to $C2$, and indeed comparing these globally over our domain is
an excellent check that the constraint equations are enforced. In
figure \ref{fig:C2C} we show $C$ and $C2$ for a black hole with
$\rho_{0} = -0.71$, and we see the difference of the functions is very
small compared to $C$, and furthermore $C$ is very small compared to
$A,B$. Thus this method, whilst appearing rather mysterious, does seem
to give very good results in practice.

We reconstruct the values of $A, B$ at the coordinate singularity
$\rho = 0$, $\chi = \pi/2$ ($C$ of course is zero), and since the
$\rho, \chi$ coordinate system is singular there, it is easiest to
consider this in the non-singular $r, z$ coordinates.  Then from the
axial symmetry at $r=0$ and the reflection symmetry at $z=\pi/2$, the
metric functions have expansions,
\begin{eqnarray}
    A(r,z) =   k_0 + k_1 r^2 + k_2 \, (\pi/2-z)^2 + O(r^4, (\pi/2-z)^4, r^2 (\pi/2-z)^2)  
\label{bc at singular point 2}
\end{eqnarray}
and similarly for $B$, and thus using the relation of $r, z$ to $\rho,
\chi$, we may simply compute how to interpolate the values at the
coordinate singularity $r = 0, z = \pi/2$ from the neighbouring points
in the $\rho, \chi$ grid.

\section{Appendix: Numerical checks}
\label{app:checks}

We find second order scaling in all physical quantities such as the
mass, temperature and entropy. However, the resolutions are
sufficiently high that increases yield very little change in the
quantities. Observe the earlier figure \ref{fig:temp_mass} where we
have plotted the thermodynamic quantities for 3 different resolutions,
which give extremely similar results where multiple resolutions may be
relaxed. This gives much confidence that the accuracy of these
solutions is high. Certainly our main conclusion is seen in this
figure, that the black holes become larger in mass and entropy than
the most non-uniform strings, and we see this is totally unaffected by
changes in resolution.

\begin{figure}
  \centerline{\psfig{file=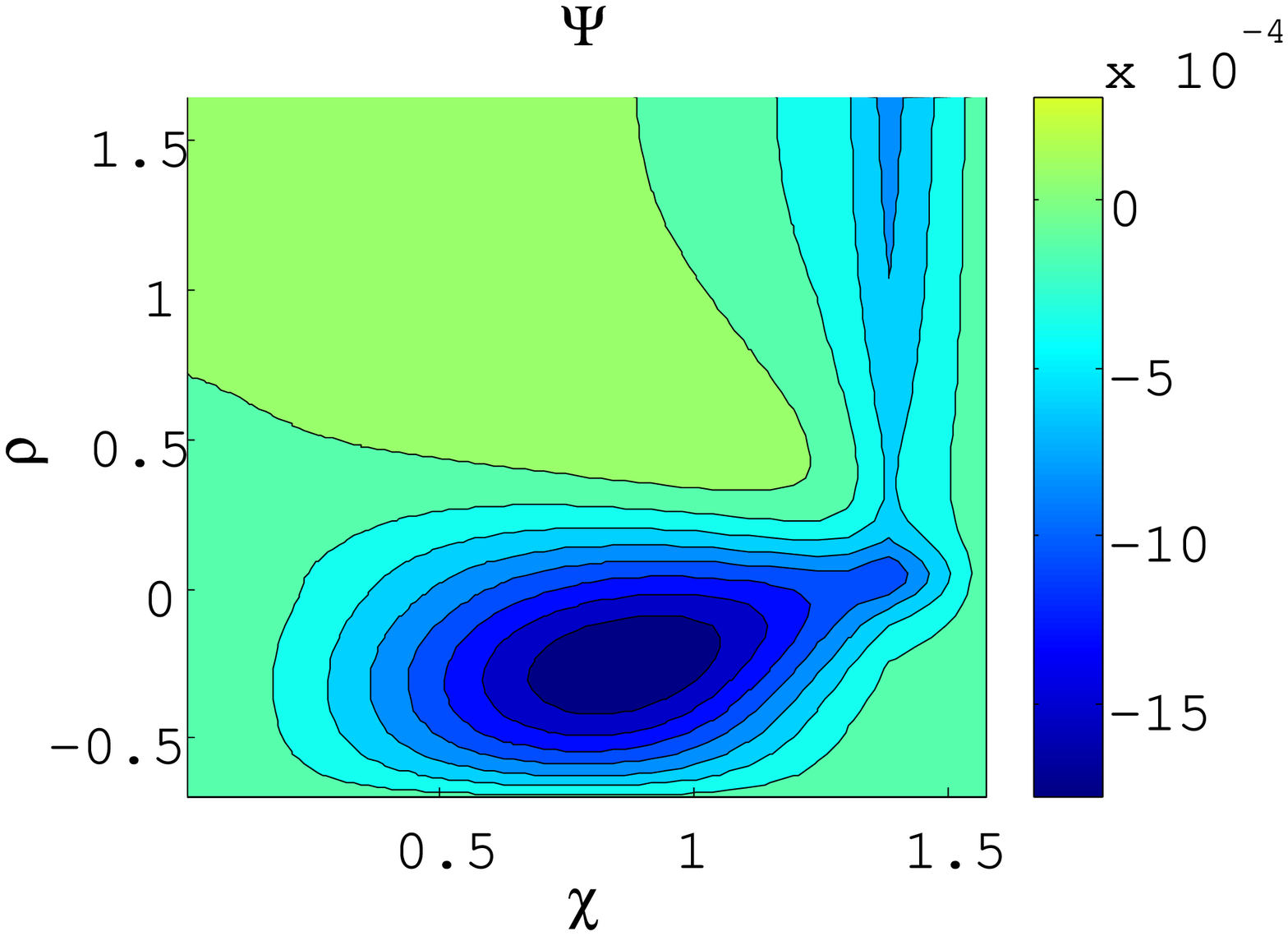,width=3.5in}
    \psfig{file=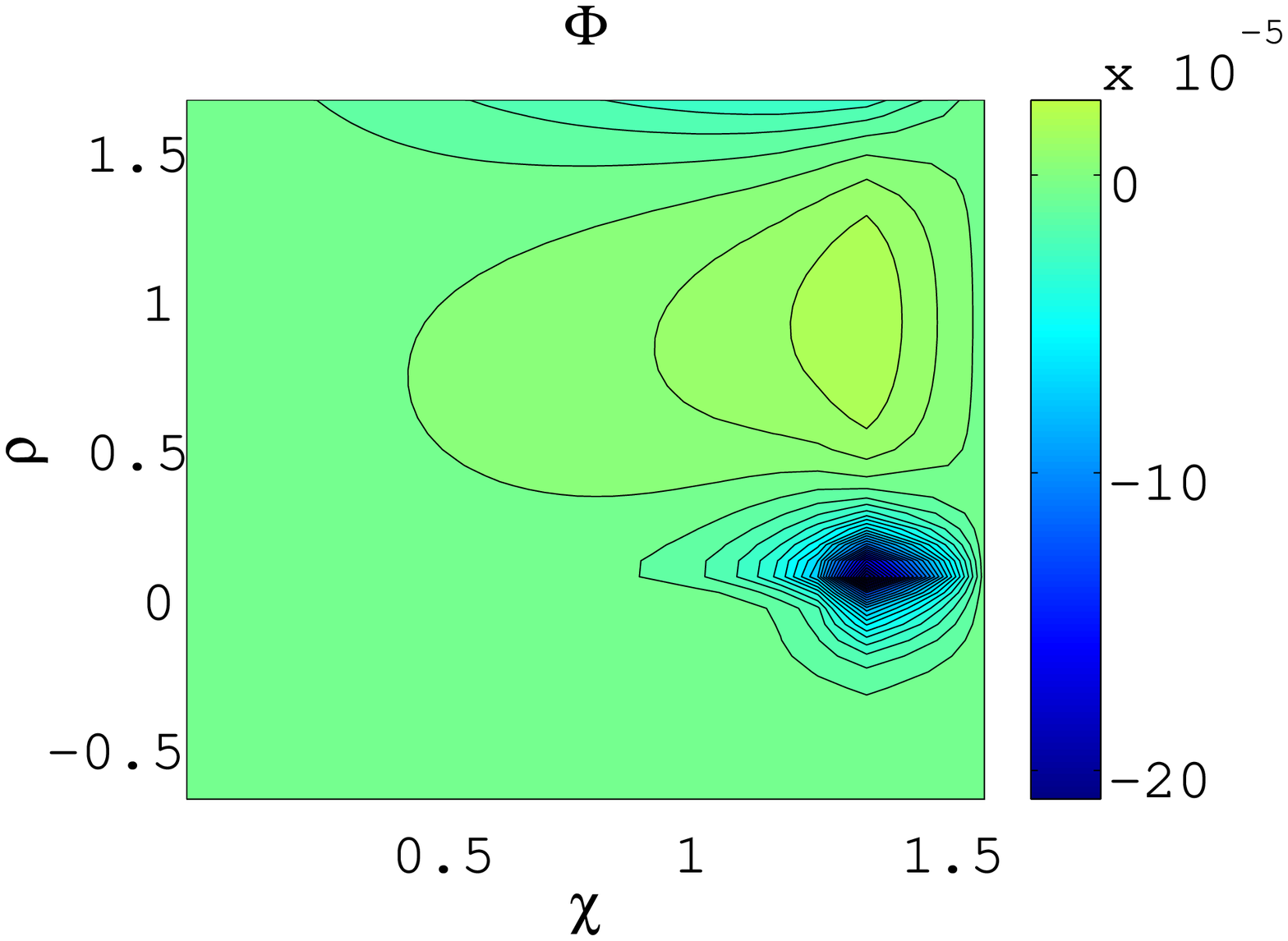,width=3.5in}}
\caption{
\label{fig:weighted_constraints}
Plots of the weighted constraint equations for $\rho_{0} = -0.71$. 
}
\end{figure}

As discussed in the main text, it is essential to explicitly test that
the constraint equations are well satisfied, as these are not imposed
directly, but only via boundary conditions and the CR relations.  For
$\rho_{0} = -0.71$ we plot in figure \ref{fig:weighted_constraints}
the weighted constraints $\Psi$ and $\Phi$. We see that they are
suitably small, rising to their maximum near the symmetry axis or
coordinate singularity. However, it is very difficult to interpret
these constraint violation values in terms of their physical effect. A
nice check that these small violations are sufficiently small that the
physics of the solution is unaffected by them is given in Table
\ref{table:constraint} where we show the average values of the
weighted constraints and $C2-C$ (which also gives a measure of how
well $\cfg$ is satisfied) over $\rho$ and $\chi$ for three resolutions
with different $\rho_0$.  As the numerical resolution is increased,
the averaged constraint violation values decrease significantly (not
quite as quickly as second order scaling, but then our discretisation
geometry is rather complicated so this would not be expected),
indicating the constraints become increasingly well satisfied, as we
would hope for.  The geometry and other properties of the solutions
varies very little as the resolution is increased, and thus the
constraint violations must be very small in terms of their physical
effect.

\begin{figure}
\centerline{\psfig{file=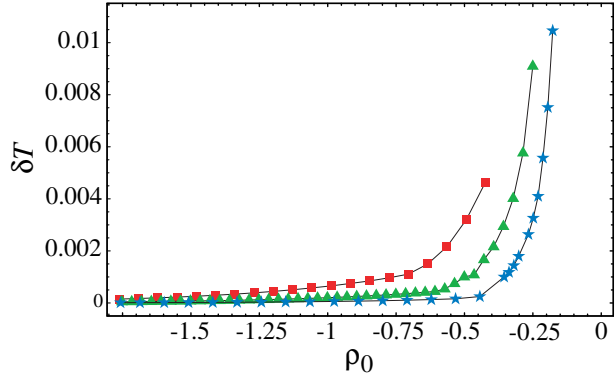,width=3.2in}  }
\caption{
  Plot showing the variation of the temperature on the horizon,
  $\delta \mathcal{T}_{error} = \left(\mathcal{T}_{max} -
    \mathcal{T}_{min}\right)/\mathcal{T}$, for three resolutions; our
  maximum 140*420 and two lower resolutions, 70*210 and
  36*106.
\label{fig:temp_error}
}
\end{figure}

In order to assess the absolute physical error in these small
constraint violations globally, we advocate comparing the values of
$C$ and $C2$ shown in the previous Appendix over the whole domain.
Since these agree extremely well, this is again excellent evidence
that $\cfg$ is effectively very well satisfied, since $C2$ is
integrated from this constraint, and $C$ is obviously derived from the
elliptic equations.  Yet another physical test we may perform is to
compute the horizon temperature as a function of $\chi$ along the
horizon. Again the $\cfg$ constraint should ensure this is constant,
yet it is the $\cffgg$ constraint that we actually impose at the
horizon for the elliptic equations.  Thus if $\cfg$ is well satisfied,
then the CR constraint structure is working well. In figure
\ref{fig:temp_error} we plot the maximum variation, $\delta
\mathcal{T}_{error} = \left(\mathcal{T}_{max} -
  \mathcal{T}_{min}\right)/\mathcal{T}$, of the temperature for
varying $\rho_{0}$ for our highest resolution, and two lower
resolutions. These variations are small, implying the constraints are
indeed well satisfied. Furthermore they decrease very nicely with
increasing resolution indicating the constraints are behaving well
numerically, and are free from systematic violations. We find maximum
variations for the largest black holes, as expected as gradients build
up near the horizon at $\chi = \pi/2$ due to the limited resolution at
the axis.  However the variation is still only $\sim 1\%$ for the
largest black hole we relaxed.

In the Table \ref{table:rhomax} we show the temperature, entropy and mass for a black hole using different $\rho_{max}$. We see these quantities (and indeed all others) hardly change, indicating our choice of $\rho_{max}$ is sufficiently large.

\begin{table}
\begin{center}
\begin{tabular}{r|cc cccc}
\hline   \hline
           & $ \left\langle |\Psi| \right\rangle $  
           & $ \left\langle |\Phi| \right\rangle $ 
           & $ \left\langle |C2-C| \right\rangle $ &
\\   \hline
$\rho_0= -1.1$  
\\ 
$140\times420$&$4.3\times10^{-5}$&$2.9\times10^{-6}$&$1.0\times10^{-5}$&\\ 
$70\times210$&$6.6\times10^{-5}$&$7.2\times10^{-6}$&$3.1\times10^{-5}$&\\ 
$35\times105$&$1.5\times10^{-4}$&$1.8\times10^{-5}$&$8.6\times10^{-5}$&   
&  
\\   \hline
$\rho_0= -0.71$  
\\ 
$140\times420$&$1.8\times10^{-4}$&$1.6\times10^{-5}$&$3.6\times10^{-5}$&  \\ 
 $70\times210$&$2.6\times10^{-4}$&$3.2\times10^{-5}$&$8.6\times10^{-5}$&  \\ 
$35\times105$&$5.1\times10^{-4}$&$5.2\times10^{-5}$&$1.7\times10^{-4}$&   
&  
\\   \hline
$\rho_0= -0.36 $
\\
$140\times420$&$7.1\times10^{-4}$&$1.0\times 10^{-4}$&$2.2\times10^{-4}$ &  \\ 
$70\times210$ &$1.2\times10^{-3}$&$2.1\times 10^{-4}$&$4.3\times10^{-4}$ 
&  \\  
$35\times105$ & --- &  --- & ---   
&  \\ 
\hline \hline   
\end{tabular}
\caption[short]{
This table shows averaged violations of the weighted constraint equations and $C2-C$ for three resolutions and three different black holes. 
The average of absolute values is taken over the whole domain.
}
\label{table:constraint}
\end{center}
\end{table}

\begin{table}
\begin{center}
\begin{tabular}{r|ccccc|lc}
\hline   \hline 
 $70\times210$     & $|\delta \mathcal{T}/\mathcal{T}|$ 
              & $|\delta\mathcal{S}/\mathcal{S}|$  
              & $|\delta M/ M |$ & 
\\   \hline
$\rho_{max}=4.6$   &$3.5 \times 10^{-4}$
      &$1.3 \times 10^{-3}$ 
      &$1.1 \times 10^{-2}$
\\ 
2.5   &$3.6 \times 10^{-3}$
      &$1.9 \times 10^{-2}$
      &$9.3 \times 10^{-2}$
\\ 
1.4   &$ 5.1\times 10^{-2}$
      &$1.8 \times 10^{-1}$
      &$4.0 \times 10^{-1}$
\\ 
\hline \hline   
\end{tabular}
\caption[short]{
This table shows the variations of the temperature, entropy and asymptotic mass for a black hole with $\rho_{0}=-0.36$ using different $\rho_{max}$. 
The variation of the quantities is defined as $\delta \mathcal{T}/\mathcal{T}= 1-  \mathcal{T}/\mathcal{T}_{\rho_{max}=5.7}$, dividing by $\mathcal{T}$ for $\rho_{max}=5.7$.
}
\label{table:rhomax}
\end{center}
\end{table}

%
\section{Appendix: Demonstration that First Law does not test the constraints}
\label{app:firstlaw}
%

In this brief Appendix we demonstrate our claim that deriving the
asymptotic mass by integrating the First Law $dM = \mathcal{T}
d\mathcal{S}$ along our branch of numerical solutions (for fixed
asymptotic radius) does not test whether the constraint equations are
satisfied. This is also true for the related Smarr law. Thus while it
is useful to check that the First Law is satisfied, as it checks the
elliptic equations are well satisfied and have boundary conditions
imposed compatibly with their regular singular behaviour, we should
not be lulled into a false sense of security. We satisfy the elliptic
equations directly in this relaxation method, and to high accuracy,
whereas the constraint equations are imposed indirectly, via the
boundary conditions. Hence these are the equations we should worry may
have numerical errors, and it is crucial to separately check these, as
was done in the previous Appendix.

The First Law can be classically derived, eg. as in \cite{WaldBH}.
Here we simply sketch the derivation, considering which components of
the Einstein tensor are involved, and therefore which can be
numerically tested by the First Law.  Consider the expression,
\begin{equation}
\tilde{S}(g_{\mu\nu}) = \int_\mathcal{M} d^6x \sqrt{-g} R(g_{\mu\nu}) 
\end{equation}
for a manifold $\mathcal{M}$ with metric $g_{\mu\nu}$. This is not the
action, since for the action we must subtract the Gibbons-Hawking term
at the boundaries $\delta \mathcal{M}$. Clearly $\tilde{S}$ (unlike
the true action) vanishes for any solution of the equations of motion,
as the Ricci scalar will always vanish locally.

The first law can be derived in our static case by considering
$g_{\mu\nu} = g^{(0)}_{\mu\nu} + \delta g_{\mu\nu}$, where
$g^{(0)}_{\mu\nu}$ is a static solution of the Einstein equations,
\emph{and} the perturbation $\delta g_{\mu\nu}$ also satisfies the
static linearised perturbation equations. Thus, $\tilde{S}$ vanishes
when evaluated on both $g^{(0)}$ and $g$. However, in the usual way we
can write,
\begin{equation}
\tilde{S}(g_{\alpha\beta}) - \tilde{S}(g^{(0)}_{\alpha\beta}) = \int_\mathcal{M} d^6x \sqrt{-g^{(0)}} G_{\mu\nu}(g^{(0)}_{\alpha\beta}) \delta g^{\mu\nu} + \left[ V_{\mu\nu} \delta g^{\mu\nu} + W_{\mu\nu} \partial_n \delta g^{\mu\nu} \right]_{\delta \mathcal{M}} + O(\delta g^2) 
\end{equation}
where $\partial_n$ is the derivative normal to the boundary, and $V,
W$ give boundary terms that arise to eliminate derivatives of $\delta
g$ in the integral term. Since the first 3 terms all vanish, we are
left simply with the boundary terms, and these give rise to the First
Law, linearised in $\delta g$, when evaluated on the horizon and
asymptotically. However, let us now consider the above in our
numerical context. If numerically we see the First is well satisfied,
does this imply all the Einstein equations, both elliptic and
constraints, are therefore well satisfied?

Naively this appears so. From the terms $\tilde{S}(g_{\mu\nu})$,
$\tilde{S}(g^{(0)}_{\mu\nu})$ we test the weighted average of the
Ricci scalars, $R(g)$ and $R(g^{(0)})$, and from the linear variation
term we test a weighted average of $G_{\mu\nu}(g^{(0)})$. However,
given that our metric \eqref{eq:simplemetric} is diagonal, and further
more, our perturbation $\delta g$ is therefore also diagonal, these
three terms then \emph{only} test the diagonal components of the
Einstein tensor.  Thus these weighted integrals simply do not involve
the $\crz$ Einstein equation (or equivalently $\cfg$). Even worse, due
to the conformal invariance of the $r, z$ (or equivalently $f, g$)
block of the metric, the perturbation is restricted so $\delta g_{rr}
= \delta g_{zz}$ and therefore the integrals also do not involve
$\crrzz$ (or equivalently $\cffgg$).

Now it is clear that if we take any solution of the elliptic
equations, $\tilde{g}^{(0)}$, totally ignoring the constraints, and
perturb this by $\delta \tilde{g}$ which again only satisfies the
static linearised elliptic equations, we can perform exactly the same
manipulations to obtain the above equation, and consequently the usual
First Law. Hence for `solutions' only obeying the elliptic equations
we would still see the First Law being well observed, even though we
had made no attempt to satisfy the constraint equations.

Thus, our specific form of the metric exactly ensures the Einstein
equations that are the `constraints' in this elliptic context, $\crz$
and $\crrzz$, do not affect the First Law, even if they are not
satisfied due to numerical error. For exactly the same reasons, the
closely related Smarr law also has no dependence on the constraint
equations, when using our metric choice, and similarly cannot provide
a numerical test of the constraints.


\newpage

\bibliography{KKpaperv2.bbl}

\end{document}